\documentclass[pra,floatfix,footinbib,reprint,superscriptaddress]{revtex4-2}
\usepackage[commandnameprefix=always]{changes}

\usepackage{graphicx}
\usepackage{dcolumn}
\usepackage{times}
\usepackage{qcircuit}
\usepackage{physics}
\usepackage{bbold}
\usepackage{amsthm, amssymb}
\usepackage{tabularx}
\usepackage{subfigure}
\usepackage{tikz}
\usepackage{dsfont}
\usepackage[normalem]{ulem}
\usetikzlibrary{matrix}
\usepackage[pdftex, colorlinks, allcolors=blue]{hyperref}
\newcommand{\swap}{\textsf{SWAP}}
\newcommand{\czeroswap}{\textsf{C}_0\textsf{SWAP}}
\newcommand{\cswap}{\textsf{CSWAP}}
\newcommand{\coneswap}{\textsf{C}_1\textsf{SWAP}}
\newcommand{\czeroz}{\textsf{C}_{0}\textsf{Z}}
\newcommand{\conez}{\textsf{C}_{1}\textsf{Z}}
\newcommand{\cz}{\mathsf{CZ}}
\newcommand{\DR}{\mathrm{DR}}
\newcommand{\SR}{\mathrm{SR}}

\DeclareSymbolFont{sfletters}{OML}{cmbrm}{m}{it}

\begin{document}

\title{Quantum random access memory architectures using 3D superconducting cavities}

\author{D.~K.~Weiss}
\email{daniel.weiss@yale.edu}
\author{Shruti Puri}
\author{S. M. Girvin}
\address{Departments of Applied Physics and Physics, Yale University, New Haven, 06511, CT, USA}
\address{Yale Quantum Institute, Yale University, New Haven, CT 06511, USA}
\date{\today}

\begin{abstract}
Quantum random access memory (QRAM) is a common architecture resource for algorithms with many proposed applications, including quantum chemistry, windowed quantum arithmetic, unstructured search, machine learning, and quantum cryptography. 
Here we propose two bucket-brigade QRAM architectures based on high-coherence superconducting resonators, which differ in their realizations of the conditional-routing operations. In the first, we directly construct cavity-controlled controlled-$\mathsf{SWAP}$ ($\cswap$) operations, while in the second we utilize the properties of giant-unidirectional emitters (GUEs).
For both architectures we analyze single-rail and dual-rail implementations of a bosonic qubit. In the single-rail encoding we can detect first-order ancilla errors, while the dual-rail encoding additionally allows for the detection of photon losses. 
For parameter regimes of interest the post-selected infidelity of a QRAM query in a dual-rail architecture is nearly an order of magnitude below that of a corresponding query in a single-rail architecture. 
These findings suggest that dual-rail encodings are particularly attractive as architectures for QRAM devices in the era before fault tolerance.

\end{abstract}

\maketitle

\section{Introduction}

Many quantum algorithms of interest presume the existence of a QRAM device capable of querying a (classical or quantum) memory in superposition. Perhaps the most well known application of QRAM is as the oracle in Grover's search algorithm~\cite{Grover1996, nielsenchuang2010}. Recently, criticisms have emerged of the utility of QRAM in the context of such ``big data" problems~\cite{Jaques2023}, particularly for algorithms like Grover that require a large number of calls to the QRAM. However, quantum algorithms utilizing QRAM where quantum advantage may still exist~\cite{Jaques2023} include modern versions of Shor's algorithm
\cite{Gidney2021, Gidney2019}, quantum chemistry algorithms~\cite{Babbush2018, Berry2019}, algorithms for solving the dihedral hidden-subgroup problem~\cite{Kuperberg2011} and the HHL algorithm~\cite{HHL}.
To get a sense of the scale of devices relevant for near-term example demonstrations of quantum advantage, the algorithm presented in Ref.~\cite{Babbush2018} for the quantum simulation of jellium in a classically intractable parameter regime  requires a modest-size QRAM with only 8 address qubits and a bus (or ``word" length) of 13 qubits.

Existing proposals for QRAM are based on quantum optics~\cite{Giovannetti2008architectures}, Rydberg atoms~\cite{Hong2012}, photonics~\cite{Chen2021} and circuit quantum acoustodynamics~\cite{Hann2019}. See Ref.~\cite{Liu2023} and references therein for a more comprehensive review. Each proposal utilizes the celebrated bucket-brigade architecture~\cite{Giovannetti2008}, which promises a degree of noise resilience as compared to more straightforward algorithmic implementations of QRAM~\cite{nielsenchuang2010}. Nevertheless, actually realizing a QRAM device appears to be extremely difficult, 
due in part to additional concerns regarding the noise-sensitivity of active vs. inactive components in a bucket-brigade QRAM~\cite{Arunachalam2015, Ciliberto2018}. Recently, Hann {\it et al.}~\cite{Hann2021} helped to address this issue, showing that the bucket-brigade architecture still enjoys a poly-logarithmic scaling of the infidelity of a QRAM query with the size of the memory even if all components are active.

In parallel to this theory work, there has been enormous experimental progress in quantum information processing using 3D superconducting cavities, including demonstrations of millisecond-scale coherence times~\cite{Milul2023, Rosenblum2018, Chakram2021}, as well as high-fidelity beamsplitter operations~\cite{Chapman2022, Lu2023}. Leveraging these results, here we propose bucket-brigade QRAM implementations based on superconducting cavities. Our proposed architectures utilize recently developed mid-circuit-error-detection and erasure-detection schemes~\cite{Tsunoda2022, Teoh2022}, boosting the query fidelity by utilizing post-selection. Among the algorithms utilizing QRAM, some may tolerate such a non-deterministic repeat-until-success procedure, as opposed to those which require interleaved QRAM calls. One example is the HHL algorithm, which can utilize QRAM for state preparation~\cite{HHL}.

The key feature of bucket-brigade QRAM that enables its relative noise insensitivity is the conditional routing of quantum information based on the states of address qubits~\cite{Giovannetti2008, Hann2021}. In this work we propose two QRAM architectures that achieve this gate primitive in distinct ways.
In the first architecture we directly construct a cavity-controlled controlled-$\mathsf{SWAP}$ ($\cswap$) gate. While an ancilla transmon is required to provide the requisite nonlinearity, we can detect and post-select away first-order transmon errors~\cite{Teoh2022, Tsunoda2022, Kubica2022}. This proposal benefits from requiring no new hardware components beyond what has already been experimentally implemented in Refs.~\cite{Lu2023, Chapman2022, Chou2023}.
Our second approach utilizes the physics of giant unidirectional emitters (GUEs)~\cite{Guimond2020, Gheeraert2020, Kannan2023} to realize the conditional routing operation, where again first-order transmon errors are detectable. We term the first the ``$\cswap$ architecture" and the second the ``GUE architecture."
In both architectures we explore single- and dual-rail~\cite{Teoh2022, Tsunoda2022} implementations. The dual-rail approach doubles the hardware cost, however it additionally allows for the first-order detection of photon loss in the cavities. This boosts the post-selected query fidelity $F$, with in particular $F>0.8$ for a QRAM with 8 address qubits in both architectures (in the remainder of this work, all query infidelities are post-selected unless noted otherwise).

Our paper is structured as follows. In Sec.~\ref{sec:bb_QRAM} we review bucket-brigade QRAM. In Sec.~\ref{sec:DR} and Sec.~\ref{sec:GUE} we explore the $\cswap$ and GUE architectures, respectively, detailing how each gate primitive is executed and computing the overall query fidelity. We discuss our results and conclude in Sec.~\ref{sec:disc}.

\section{Bucket-Brigade QRAM}
\label{sec:bb_QRAM}

\begin{figure*}
    \centering
    \includegraphics[width=2\columnwidth]{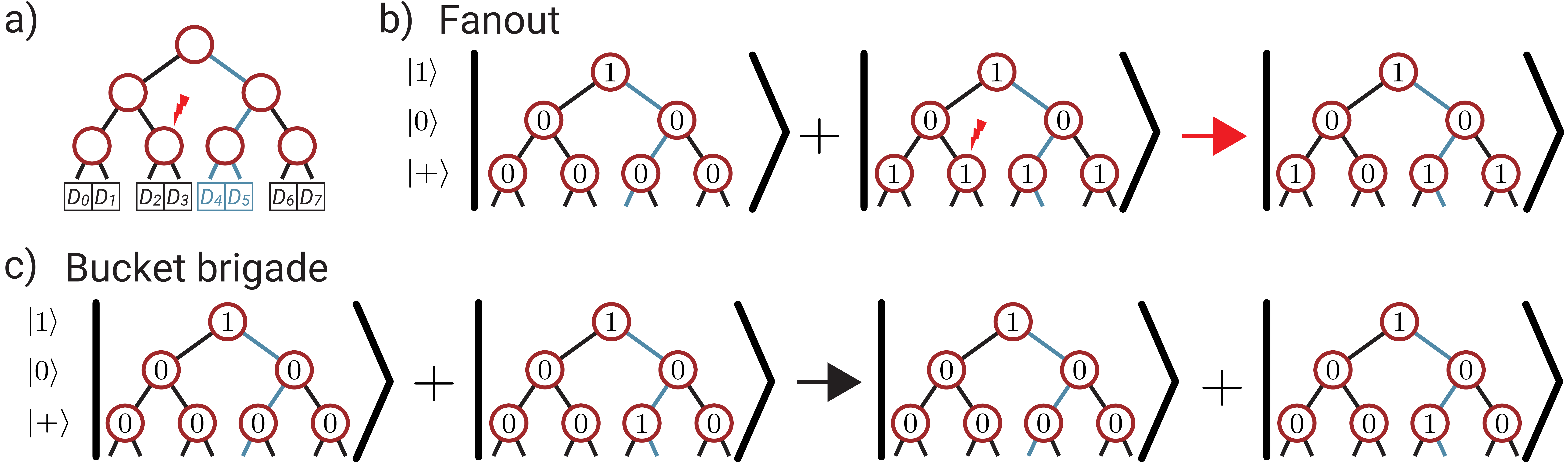}
    \caption{\label{fig:fanout_bb} {Comparison of fanout and bucket-brigade QRAM with two-level routers. a) By constructing a tree of quantum routers, one can access data in superposition. We consider the state of the tree in a query to $D_{4}$ and $D_{5}$ (highlighted) where the indicated router experiences an amplitude-damping error (the general argument holds also for other error types, see Ref.~\cite{Hann2021}). b) In a fanout approach \cite{nielsenchuang2010}, all routers are set at each level according to the state of the address at that level, leading to a GHZ-like state. The resulting state of the QRAM tree is thus highly entangled and susceptible to decoherence. An amplitude-damping event on the indicated router causes the state of the tree to collapse and the query to fail.
    c) In the case of bucket-brigade QRAM \cite{Giovannetti2008, Hann2021}, only routers participating in a query to a given branch are activated. This leads to a less entangled state of the QRAM tree as compared to fanout QRAM. Amplitude damping now can not occur on the router indicated in a) and the superposition state is not disrupted. }
    }
\end{figure*}
The purpose of a QRAM device is to realize the unitary operation
\begin{align}
\label{eq:QRAM}
    \sum_{i}\alpha_{i}|i\rangle|0\rangle \rightarrow \sum_{i}\alpha_{i}|i\rangle|D_{i}\rangle,
\end{align}
where classical data $D_{i}$ specified by address $|i\rangle$ is accessed in superposition and stored in the state of the bus. (In this work we take the memory to be a classical database, however in general the memory may be quantum, see e.g., Refs.~\cite{Ambainis2014, Arunachalam2015, Jaques2023}).
The first and second registers in Eq.~\eqref{eq:QRAM} are the $n$ address qubits and bus qubit, respectively, where we query a database of $N=2^n$ classical bits.

{To appreciate the benefits of the bucket-brigade approach to QRAM, it is useful first to review so-called ``fanout" QRAM \cite{nielsenchuang2010}. Quantum routers are arranged in a tree-like structure, see Fig.~\ref{fig:fanout_bb}(a). The states of the routers at the $\ell^\text{th}$ level of the tree are set by the $\ell^\text{th}$ address qubit. The bus qubit is then routed into the tree, and arrives at the memory locations specified by the address qubits. After copying the classical data into the bus and routing it back out, the router-state setting operation is performed again to uncompute and disentangle the address and bus qubits from the routing tree, thus achieving the operation Eq.~\eqref{eq:QRAM}. 
}

{The issue with this proposal is the high susceptibility to decoherence \cite{Hann2021}. Each layer of CNOTs effectively creates a maximally-entangled Greenberger-Horne-Zeilinger state among the routers at each level, see Fig.~\ref{fig:fanout_bb}(b). If any one router decoheres, this collapses the superposition at that level and reduces the query fidelity on average by a factor of 2 \cite{Giovannetti2008}. This is clearly not a scalable approach, as there are exponentially many (in $n$) quantum routers in the bottom layers of the tree. }

{The bucket-brigade approach aims to achieve the same operation \eqref{eq:QRAM} as fanout QRAM (and indeed also uses a tree structure of quantum routers) but with an algorithm that attempts to minimize as much as possible entanglement between different routers \cite{Giovannetti2008}. The address qubits (followed by the bus) are fed into the top of the tree one-by-one, and are conditionally routed based on the states of previous address qubits. The resulting state of the routing tree (after all address qubits have been routed in) is such that routers at the same level of the tree are not entangled with one another. Thus, an error on one router does not cause a catastrophic collapse of the state of the QRAM tree, see Fig.~\ref{fig:fanout_bb}(c). It instead only disrupts queries along branches passing through the decohered router. }

{The original bucket-brigade proposal called for three-level routers, with states $|W\rangle,|0\rangle,|1\rangle$ \cite{Giovannetti2008}. The inactive ``wait" state $|W\rangle$ acts trivially (identity) for all conditional routing operations~\cite{Giovannetti2008, Hann2021}. It was originally thought \cite{Giovannetti2008, Arunachalam2015, Ciliberto2018} that the favorable noise properties of bucket-brigade QRAM were predicated on the wait state being decoherence free. However, it was realized in Ref.~\cite{Hann2021} that the three-level bucket-brigade QRAM is noise resilient even if the wait state is subject to decoherence. This resiliency is due to the limited entanglement among branches as well as the restricted propagation of errors between branches. Roughly speaking, errors in one branch do not eventually poison (via the conditional routing operations) queries in other branches, due to the trivial action of the conditional routing operations on the $|W\rangle$ state. Hann {\it et al.}
prove in Ref.~\cite{Hann2021} that the query infidelity for a QRAM with three-level routers is bounded by
\begin{align}
\label{eq:infidel_3}
1-F\leq A\epsilon N_{\mathrm{ts}} \log_{2}(N),
\end{align}
where $A\approx 4$, $\epsilon$ is the error probability per time step and $N_{\rm ts}$ is the number of timesteps in a QRAM query (which we compute below).} 

{Perhaps surprisingly, the noise resilience persists even if two-level routers are utilized, see Fig.~\ref{fig:fanout_bb}. Errors can propagate more freely than in the three-level router case, as conditional routing operations conditioned on router state $|0\rangle$ can cause errors in one branch to interfere with queries to otherwise decoherence-free branches (assuming that the tree is initialized in the vacuum state). However, there is still limited entanglement in the tree and only some errors are allowed to propagate. In this case the query infidelity is \cite{Hann2021}
\begin{align}
\label{eq:fidel_two_level_router}
1-F\leq A\epsilon N_{\mathrm{ts}} \log_{2}(N)[1+\log_{2}(N)],
\end{align}
which is still poly-logarithmic in the size of the memory. In this work we consider architectures with two-level routers. We thus utilize Eq.~\eqref{eq:fidel_two_level_router} when calculating query infidelities and proceed by calculating the error-per-timestep $\epsilon$. 
}

To realize a bucket-brigade QRAM, three primitive operations are required: (i) setting the state of a router, (ii) conditional routing and (iii) copying classical data into the state of the bus. We detail how each of these gate primitives is executed for each of our proposals below, and analyze the resulting query fidelities.

\section{$\cswap$ architecture}
\label{sec:DR}

In the $\cswap$ architecture, quantum information is stored in high-Q superconducting memories that are coupled via beamsplitter elements~\cite{Chapman2022, Lu2023, Gao2018, Gao2019}. In our protocol, the comparatively low-coherence transmons are only excited briefly during gate operations and are disentangled from the cavities at the conclusion of each gate. Moreover, the transmons are first-order error detected using the techniques described in Refs.~\cite{Tsunoda2022, Teoh2022}.
\begin{figure*}
    \centering
    \includegraphics[width=2\columnwidth]{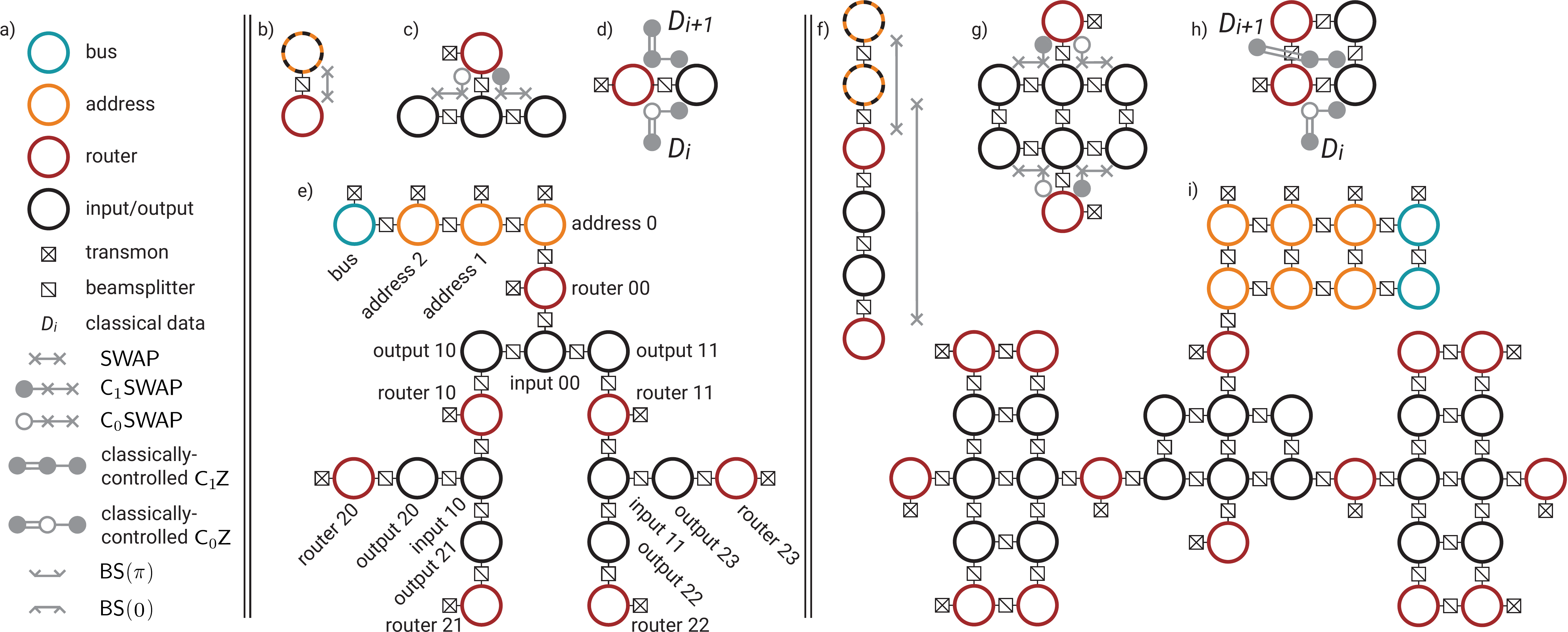}
    \caption{\label{fig:DR_3_address} 
    Single- and dual-rail $\cswap$ QRAM. (a) We include a legend explaining our coloring scheme, the notation for hardware elements, and gate schematics for this figure as well as later figures. For the single-rail (b-e) and dual-rail (f-i) implementations, we detail the gates required for (b, f) setting the state of the router, (c, g) performing the $\czeroswap$ and $\coneswap$ operations and (d, h) copying classical data into the state of the bus. The cavities outlined in dashed orange/black indicate that setting the state of the router occurs both at the top of the tree (involving the address cavity) as well as at intermediate nodes (involving output cavities). 
    We show example hardware layouts for 3-bit (e) single-rail and (i) dual-rail QRAM. The full quantum circuit associated with a QRAM query on the hardware of (e) is shown in Appendix~\ref{app:qcircuit}. The naming scheme for routers and input/output elements is e.g. router ij, with i corresponding to the layer and j to the position going from left to right. We emphasize that in our proposed bucket-brigade QRAM schemes the classical data is not stored in any quantum elements before it is copied into the state of the bus. The data instead determines which quantum gates to perform. Thus instead of ``memory cells" at the bottom of the tree (which would require additional, expensive quantum hardware) we have routers which allow access to data stored at two distinct locations, see (d, h).
    }
\end{figure*}

\subsection{Gate protocol}

Setting the state of a router is performed straightforwardly utilizing $\mathsf{SWAP}$ operations, see Fig.~\ref{fig:DR_3_address}(b). This operation is performed at the top of the tree as well as between output cavities and their neighboring routers, see Fig.~\ref{fig:DR_3_address}(e). 

Conditional routing in this architecture is achieved by direct construction of $\cswap$ operations. It is important to emphasize that in our construction, cavities serve as the controls for conditional routing. This is to be contrasted with previous work that realized a $\cswap$ operation using a transmon as the control~\cite{Gao2019}. Here, transmons are used only as ancillas.

The $\textsf{CSWAP}$ gate is enabled by the
joint-parity gate~\cite{Teoh2022, Tsunoda2022, Gao2019}  
\begin{align}
\mathsf{JP}_{ab} = |g\rangle\langle g|  + \exp(i\pi[\hat a^{\dagger} \hat a + \hat b^{\dagger} \hat b])|f\rangle\langle f|.
\end{align}
where $\hat a$ and $\hat b$ are the annihilation operators of the two cavities referred to as $a$ and $b$, while $|g\rangle$, $|f\rangle$ are the two computational states of the transmon. Importantly, the excited state $|e\rangle$ is reserved for first-order error detection of transmon decay~\cite{Teoh2022, Tsunoda2022, Kubica2022}. We take the transmon to be coupled to cavity $a$ with dispersive Hamiltonian $\frac{\chi}{2}\hat{a}^{\dagger}\hat{a}\hat{\sigma}_{z}$, where Pauli matrices are defined in the $|g\rangle,|f\rangle$ manifold and $\chi$ denotes the strength of the dispersive interaction.
The joint-parity gate is performed by first exciting the ancilla via a Hadamard gate, activating a beamsplitter between cavities $a$ and $b$ for time $2\pi/\chi$, and then applying another Hadamard to the ancilla, see Refs.~\cite{Teoh2022, Tsunoda2022} for further details.
Two applications of this gate separated by rotations on the ancilla transmon realizes an entangling operation between the two cavities~\cite{Teoh2022, Tsunoda2022}
\begin{align}
\mathsf{ZZ}_{ab}(\theta) &= e^{i\frac{\pi}{4}\hat\sigma_{y}}\mathsf{JP}_{{ab}} e^{-i\frac{\theta}{2}\hat\sigma_{x}}\mathsf{JP}_{{ab}}e^{-i\frac{\pi}{4}\hat\sigma_{y}} \\ \nonumber 
&= \cos\frac{\theta}{2}\hat{\mathds{1}} -i\sin\frac{\theta}{2}e^{i\pi[\hat a^{\dagger} \hat a + \hat b^{\dagger} \hat b]}\hat\sigma_{z},
\end{align}
If the transmon is detected in $|e\rangle$ or $|f\rangle$ at the conclusion of the gate, we infer that an ancilla decay or dephasing event has occurred, respectively.
In the following, we assume that the transmon always begins in its ground state.
The effect of this gate on states in the Fock basis is then
\begin{align}
\mathsf{ZZ}_{ab}(\theta)|n_{a}n_{b}\rangle = e^{-i\frac{\theta}{2}P_{n_{a}n_{b}}}|n_{a}n_{b}\rangle,
\end{align}
where $P_{ij}=1,-1$ if the joint parity of the $a$ and $b$ modes is even or odd, respectively.
For $\theta=\pi/2$, this gate becomes a $\cz_{ab}$ operation in the computational subspace, up to the single-cavity operations $e^{-i\frac{\pi}{2}\hat a^{\dagger}\hat a}e^{-i\frac{\pi}{2}\hat b^{\dagger}\hat b}$ and a global phase~\cite{Schuch2003, Teoh2022, Tsunoda2022}
\begin{align}
\label{eq:CZ}
\cz_{ab} |n_{a} n_{b}\rangle &\equiv e^{i\pi/4}e^{-i\frac{\pi}{2}(\hat a^{\dagger}\hat a+\hat b^{\dagger}\hat b)}\mathsf{ZZ}_{ab}(\pi/2)|n_{a} n_{b}\rangle  \\ \nonumber 
&=e^{-i\frac{\pi}{2}\left[\frac{P_{n_{a}n_{b}}}{2}+n_{a}+n_{b}-\frac{1}{2}\right]}|n_{a}n_{b}\rangle.
\end{align}
To utilize this gate towards construction of a $\textsf{CSWAP}$ operation, we take inspiration from the canonical construction of $\textsf{CSWAP}$~\cite{nielsenchuang2010, Chuang1995}. This construction sandwiches a Kerr interaction between two cavities $a$ and $b$ by 50:50 beamsplitters between cavity $b$ and a third cavity $c$. The Kerr interaction (synthesized here via the $\cz_{ab}$ operation) acts as a phase shifter that either completes or undoes the beamsplitter. In this way, we obtain both a $\czeroswap$ and a $\coneswap$
\begin{align}
\czeroswap_{abc} &= e^{i\pi \hat b^{\dagger}\hat b}\mathsf{BS}_{bc}(\pi/2)\cz_{ab}\mathsf{BS}_{bc}(\pi/2), \\ \nonumber 
\coneswap_{abc} &=\mathsf{BS}_{bc}^{\dagger}(\pi/2)\cz_{ab}\mathsf{BS}_{bc}(\pi/2),
\end{align}
where $\czeroswap$ and $\coneswap$ indicate the $\mathsf{SWAP}$ is executed if the control is set to $|0\rangle$ or $|1\rangle$, respectively. In this way we achieve the conditional routing of quantum information, see Fig.~\ref{fig:DR_3_address}(c). We have defined
\begin{align}
\mathsf{BS}_{bc}(\alpha)=\exp(-i\frac{\pi}{4}[e^{i\alpha}\hat c^{\dagger}\hat b+e^{-i\alpha}\hat b^{\dagger}\hat c]),
\end{align}
thus
\begin{align}
\mathsf{BS}_{bc}(\pi/2)=\exp(\frac{\pi}{4}[\hat c^{\dagger}\hat b-\hat b^{\dagger}\hat c]).
\end{align}
The additional single-cavity rotation in $\czeroswap$ is to correct for unwanted phases. See Appendix~\ref{app:DR_CSWAP} for an explicit verification that the gate functions as intended on the states of interest $|n_{a}00\rangle, |n_{a}01\rangle, |n_{a}10\rangle, n_{a}=0,1$, where the third index refers to occupation in mode $c$.
We note that these $\textsf{CSWAP}$ gates do not behave as expected if modes $b$ and $c$ are {\em both} initially occupied. We might expect that the overall operation should be trivial. However, due both to the Hong-Ou-Mandel (HOM) effect~\cite{HOM} and to phase shifts due to the $\cz_{ab}$ gate, population is transferred out of the computational subspace. We stress that in the absence of thermal photons, we never expect these modes to be simultaneously occupied in the course of a QRAM query. Thus all population should ideally remain in the computational subspace~\footnote{We are assuming that the QRAM device is initialized in the vacuum state, as opposed to some proposed QRAM architectures that can be initialized in an arbitrary state~\cite{Hann2021}}. Nevertheless, both modes may become occupied due to thermal excitations, thus we discuss this case in Appendix~\ref{app:DR_CSWAP}.

These operations thus realize logical $\textsf{CSWAP}$s in the single-rail case. 
In the dual-rail case the logical states are given in terms of the cavity Fock states as $|0_{\mathrm{L}}\rangle = |10\rangle, |1_{\mathrm{L}}\rangle = |01\rangle$. 
Logical $\textsf{CSWAP}$ operations in the dual-rail case are then realized by performing $\czeroswap$s and $\coneswap$s on both halves of the dual-rail qubits, see Fig.~\ref{fig:DR_3_address}(g) (taking the cavity with the first index to be the top cavity). The two physical gates making up a single logical $\czeroswap$ may be performed in parallel, thus the query time is not increased by utilizing a dual-rail architecture. We note here that in the dual-rail case, it is necessary to initialize the QRAM tree in the vacuum state, as opposed to e.g. initializing each dual rail in $|0_{\mathrm{L}}\rangle$ (in the single-rail case we also initialize each cavity in vacuum, however those are logical states). This initialization choice is due again to the HOM effect discussed above. Initialization of all dual rails in $|0_{\mathrm{L}}\rangle$ would lead to cases where both target cavities of a $\textsf{CSWAP}$ are occupied, leading to leakage out of the logical subspace.

The final gate-primitive necessary is the operation for copying classical data into the bus. Following Ref.~\cite{Hann2021}, we route the bus into the tree in the state $|+_{\mathrm{L}}\rangle$. In the single-rail case this is the state $(|0\rangle+|1\rangle)/\sqrt{2}$, while in the dual-rail case this is the state $(|10\rangle+|01\rangle)/\sqrt{2}$. Once the bus reaches the bottom layer of the tree, we perform classically-controlled $\czeroz$ and $\conez$ operations between the bus and the router as shown in Fig.~\ref{fig:DR_3_address}(d), where the subscript refers to conditioning on the state of the router. By definition, we have $\conez\equiv\cz$. The $\czeroz$ gate is compiled as a $\cz$ gate followed by a $\mathsf{Z}$ gate on the bus (performed in software), such that overall a $\mathsf{Z}$ gate is applied to the bus only if the router is in the state $|0_{\rm L}\rangle$. The additional layer of classical control simply means that we execute the $\czeroz,\conez$ gates only if the classical data is set to 1. The gate operations yield in both the single- and dual-rail cases [see Fig.~\ref{fig:DR_3_address}(d,h)]
\begin{align}
[|0_{\mathrm{L}}\rangle + |1_{\mathrm{L}}\rangle][ \alpha |0_{\mathrm{L}}\rangle + \beta |1_{\mathrm{L}}\rangle ] \rightarrow
\alpha [|0_{\mathrm{L}}\rangle + (-1)^{D_{i}}|1_{\mathrm{L}}\rangle]|0_{\mathrm{L}}\rangle \\ \nonumber 
\quad + \beta [|0_{\mathrm{L}}\rangle + (-1)^{D_{i+1}}|1_{\mathrm{L}}\rangle]|1_{\mathrm{L}}\rangle,
\end{align}
ordering the states as bus, router and where $|i_{\mathrm{L}}\rangle\equiv|i\rangle$ in the single-rail case. Here, $D_{i}, D_{i+1}$ are the classical data accessed by the router at the bottom of the tree.
This scheme satisfies the ``no-extra copying" condition~\cite{Hann2021}, where the action of the gate is trivial for any locations where the bus has not been sent.

We emphasize that the error detection of ancilla errors is performed after every $\cz, \cswap$ gate, see Appendix~\ref{app:CZ_fidelity}. No ``which path" information is revealed by performing measurements on the ancillas, as they are disentangled from the tree after every gate and do not collapse the superposition of queries. This is to be contrasted with error detection of cavity photon-loss events in the dual-rail case, which must wait until the address and bus qubits have been routed out of the tree. We cannot perform mid-circuit cavity-photon-loss detection because we initialize the tree in the vacuum state, as opposed to dual-rail logical states. Thus, such error detection would reveal which path information and collapse the superposition state of the tree. For example, detecting that a photon \emph{is present} in a router (indicating occupation of a dual-rail logical state) destroys those terms in the superposition state of the tree where this router was in vacuum. If we hypothetically had access to a $\cswap$ operation that performs as expected when both target modes are occupied, then we may initialize the tree in logical dual-rail states and perform mid circuit error detection of cavity photon loss. Indeed, detecting cavity photon-loss errors becomes important in the large $n$ limit, when query times approach cavity lifetimes $T_{1}^{c}$. Thus, constructing such a $\cswap$ operation is an interesting avenue of further research.

\subsection{Resource estimates}

We estimate the hardware cost of our proposed implementations by counting the number of cavities, as the number of required beamsplitter elements and ancilla transmons scale proportionally.  We require
\begin{align}
N_{\mathrm{cav}}^{\SR} = \frac{5}{2}N+\log_{2}(N)-3, \quad 
N_{\mathrm{cav}}^{\DR} = 2 N_{\mathrm{cav}}^{\SR},
\end{align}
cavities in the single-rail and dual-rail case, respectively. $N$ is the size of the memory and is related to the number $n$ of address qubits by $N=2^n$. Here and in the following, unless otherwise stated we assume $n\geq2, N\geq4$.

We now turn to estimating the number of gates. Specifically, we estimate the number of $\cz$ gates, which underlies both the $\textsf{CSWAP}$ and the data-copying operations. We ignore the number of required $\mathsf{SWAP}$ (or 50:50 beamsplitter) operations, as these gates are fast compared to the $\cz$ gate execution time: a $\mathsf{SWAP}$ can be executed in as little as 50-100 ns~\cite{Chapman2022, Lu2023}, while a $\cz$ gate takes time $4\pi/\chi$ (ignoring the single-qubit-gate times on the transmon)~\cite{Teoh2022, Tsunoda2022}. The dispersive coupling strength $\chi$ between the cavity and ancilla transmon is typically on the order of one to a few MHz~\cite{Koch2007}, taken here and in the following to be $\chi/2\pi=2$ MHz, corresponding to a $\cz$ gate time of $t_{\cz}=1\mu$s. In the single-rail case, the total number of required $\cz$ gates is
\begin{align}
N_{\mathrm{gates}} &= 7 N - 4 \log_{2}(N) - 8.
\end{align}
In the dual-rail case, the $\cz$ operations necessary for the logical $\textsf{CSWAP}$s are executed in parallel. Thus counting the logical operation as a single gate, the total number of logical $\cz$ operations is the same as in the single-rail case [counting the data-copy operations also as logical $\cz$s, though only a single physical $\cz$ is required, see Fig.~\ref{fig:DR_3_address}(h)].

While there are $\mathcal{O}(N)$ gates, the circuit depth scales only as $\mathcal{O}(\log_{2}[N])$. This is because many of the gates may be executed in parallel: all of the gates at a single horizontal layer of the QRAM may be executed simultaneously. Moreover, we may utilize {\it address pipelining}~\cite{Jaques2023, Xu2023}; once an address qubit has been routed past the first set of output ports, the next address (or bus) qubit may be routed in. Thus the total number of time steps scales only logarithmically in $N$
\begin{align}
\label{eq:Nts}
N_{\mathrm{ts}} &= 4 (1 + 2 + 3 (n-3) + 2) + 2, \quad n\geq 3,
\\ \nonumber &= 12 \log_2(N) - 14, 
\end{align}
and $N_{\mathrm{ts}}=10$ for $n=2$. This number is the same for both single- and dual-rail.
The factor of four in Eq.~\eqref{eq:Nts} is two factors of 2 coming from the need to route qubits both in and out, and the need to perform $\czeroswap$ as well as $\coneswap$. The second address qubit needs to traverse one level of the tree and the third address needs to traverse two levels. Address qubits routed further down the tree may be pipelined, and each introduces only a constant factor of additional time steps. The final factor inside of the parentheses accounts for routing the bus, while the final factor in Eq.~\eqref{eq:Nts} corresponds to the data copying steps.

The time to complete a QRAM query is then $t_{n} = t_{\cz}N_{\rm ts}$. For the parameters considered in this work, $t_{n}$ is generally short when compared with $T_{1}^{c}$. For instance, for $n=12$ and $N=2^{12}$, we obtain $t_{12}=130\mu$s.

\subsection{{Infidelity and error mitigation by post selection}}

We now turn to estimating the overall infidelity and no-flag probability (probability of detecting no errors) of a QRAM query. 
In both the dual-rail and single-rail cases, we utilize the infidelity formula for two-level routers (as opposed to that for three-level routers). This is clear for the single-rail case (as we only utilize Fock states $|0\rangle$, $|1\rangle$), however for the dual-rail case it might appear at first glance that we have indeed implemented a three-level router [and thus can utilize the more favorable infidelity formula~\eqref{eq:infidel_3}]. The physical states $|10\rangle$ and $|01\rangle$ encode logical $|0_{\mathrm{L}}\rangle$ and $|1_{\mathrm{L}}\rangle$, respectively and the state $|00\rangle$ could then play the role of the logical wait state $|W_{\mathrm{L}}\rangle$~\cite{Giovannetti2008, Hann2021}. 
This reasoning is incorrect because we construct the logical $\textsf{CSWAP}$ by utilizing physical $\czeroswap$ and $\coneswap$ operations. 
On the one hand, errors in target cavities subject to the $\coneswap$ do indeed get stuck if the router is in the state $|00\rangle$.
On the other hand, errors in target cavities subject to the $\czeroswap$ can propagate up the tree with the router in the state $|00\rangle$. Thus, we expect the infidelity scaling for two-level routers of Ref.~\cite{Hann2021} to apply to the dual-rail case as well \footnote{The arguments of Ref.~\cite{Hann2021} with regards to two-level routers do not exactly apply, as the reasoning there was based on the ability of errors to only propagate up through routers on the left. They can be made to apply by considering separately each half of the dual rail, and noting that each of these halves is occupied approximately half the time.}.
 
\begin{figure}
    \centering
    \includegraphics[width=\columnwidth]{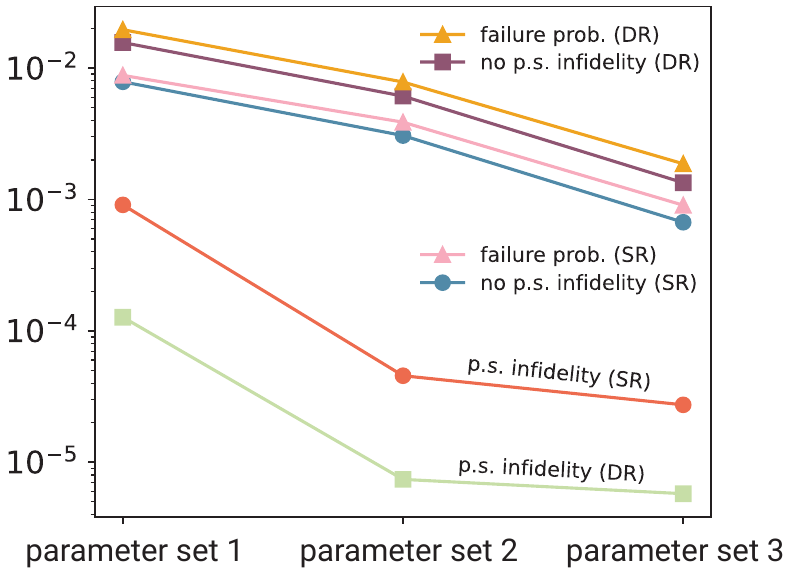}
    \caption{\label{fig:infidel_vs_PS} Infidelity, post-selected (p.s.) infidelity and failure probability of the logical $\cz$ operation as a function of parameter set for both single rail (SR) and dual rail (DR). Utilizing post-selection and parameter sets 2-3, infidelities on the order of $10^{-5}, 10^{-6}$ are possible for single and dual rail, respectively.   }
\end{figure}

We take the probability of error per time step to be $\epsilon=1-F_{g}(\cz)$~\cite{Hann2021, Chen2021}, consistent with neglecting sources of error due to beamsplitter operations. We have defined $F_{g}(\cz)$, the gate fidelity of the $\cz$ gate. We expect $F_{g}(\cz)$ to be limited by decoherence and measurement errors (when utilizing postselection), as in the absence of these non-idealities Eq.~\eqref{eq:CZ} realizes a perfect $\cz$ gate. We calculate $F_{g}(\cz), \epsilon$ both with and without post-selection on first-order errors [see Appendix~\ref{app:CZ_fidelity} for details] using three sets of parameters, see Tab. ~\ref{tab:params} and Fig.~\ref{fig:infidel_vs_PS}. Parameter set 1 (PS1) is based on values recently reported in the literature for a combined transmon and 3D resonator package~\cite{Sivak2023}.
For parameter set 2 (PS2) we use state-of-the-art coherence times for transmons~\cite{Place2021, Wang2022} and 3D resonators~\cite{Milul2023}.
In parameter set 3 (PS3) we utilize the same cavity coherence times as PS2, but make more optimistic assumptions for the transmon coherence times. We additionally include the detrimental effects of measurement errors when calculating $F_{g}(\cz),\epsilon$. Such effects occur at second order and thus at the same order as e.g. two uncaught transmon errors. They are second order because first an error must occur, then it must be misidentified.  
See Appendix~\ref{app:CZ_fidelity} for more details.

Without utilizing post-selection on first-order transmon errors, the gate infidelity is on the order of $\epsilon\sim 10^{-2}-10^{-3}$ for both single and dual rail [see Fig.~\ref{fig:infidel_vs_PS}].
However, post-selected infidelities on the order of $10^{-5}, 10^{-6}$ are possible for single rail and dual rail, respectively, utilizing estimates from parameter sets 2 and 3.

\begin{table}
\caption{\label{tab:params}Parameters used in numerical simulations and associated jump operators. The decay and dephasing rates of the cavities are $\Gamma_{1}^{c}=1/T_{1}^{c}$, $\Gamma_{\phi}^{c}=1/T_{\phi}^{c}$, respectively, implicitly defining the associated coherence times. For the transmon, decay and dephasing rates are generally state dependent \cite{Morvan2021}. We make the usual assumption of bosonic enhancement these rates \cite{Tsunoda2022}, taking $\Gamma_{\phi}^{\, t, gg}=0, \Gamma_{\phi}^{\, t, ff}=4 \Gamma_{\phi}^{\, t, ee}$ and $\Gamma_{1}^{\,t, ef} = 2\Gamma_{1}^{\,t, ge}$. The associated coherence times are defined as e.g. $T_{1}^{\,t, ef}=1/\Gamma_{1}^{\,t, ef}$. Finally, we assume a thermal population of $n_{\mathrm{th}}=0.01$.} 
\begin{ruledtabular}
\renewcommand{\arraystretch}{1.0} 
\begin{tabular}{ccccc}
& jump op. & PS1~\cite{Sivak2023} & PS2 & PS3 \\
\hline
$T_{1}^{\,c}$ & $\hat{a}$ & 0.6 ms & 25 ms~\cite{Milul2023} & 25 ms \\
$T_{\phi}^{\,c}$ & $\hat{a}^{\dagger}\hat{a}$ & 5 ms  & 106 ms~\cite{Milul2023} & 106 ms\\
$T_{1}^{\,t, ge}$ & $|g\rangle\langle e|$ & 0.2 ms & 0.5 ms~\cite{Wang2022} & 2 ms \\
$T_{\phi}^{\,t, ee}$ & $|e\rangle\langle e|$ & 0.4 ms  & 0.9 ms~\cite{Wang2022} & 4 ms \\
\end{tabular}
\end{ruledtabular}
\end{table}

\begin{figure}
    \centering
    \includegraphics[width=\columnwidth]{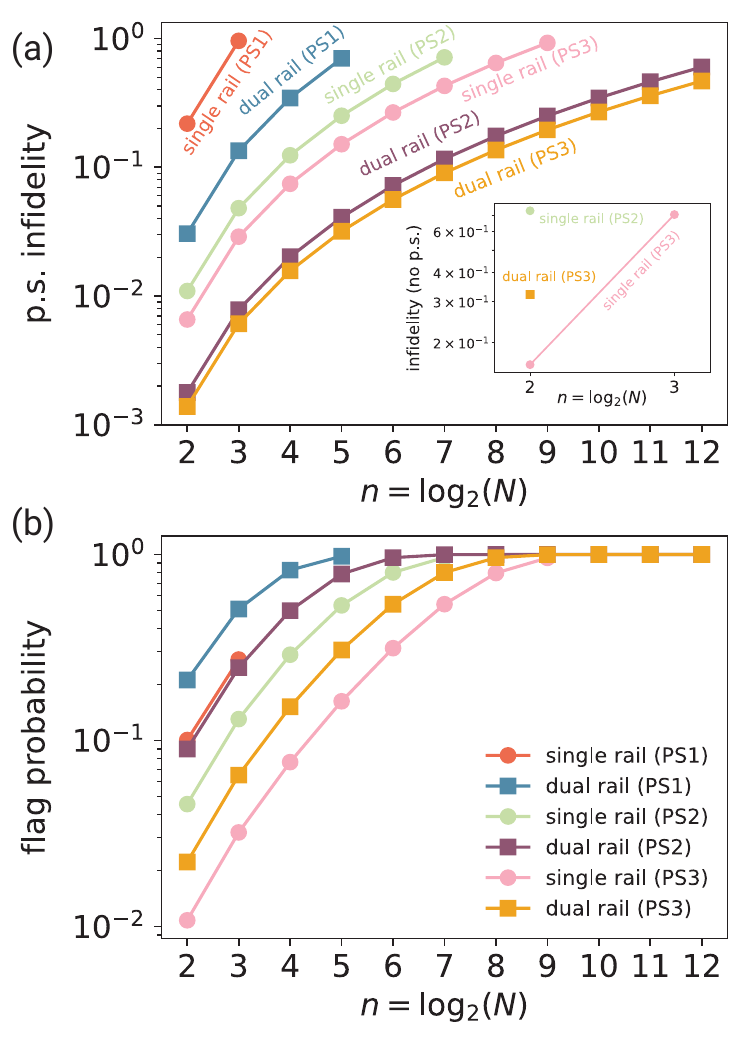}
    \caption{\label{fig:fidel_and_success} Overall infidelity and flag probability of a QRAM query as a function of $n$. 
    (a) For the three parameter sets shown in Tab.~\ref{tab:params}, we calculate the post-selected infidelity for both the single-rail and dual-rail architectures (showing only values for which the fidelity is nonzero). In the inset we plot the fidelity without utilizing post-selection. 
    The dual-rail implementation enjoys a decrease in the post-selected infidelity with respect to the single-rail case. (b) This comes at the price of increasing the flag probability.
}
\end{figure}

These low-error rates enable high fidelity QRAM queries, see Fig.~\ref{fig:fidel_and_success}(a). Comparing dual rail and single rail, a dual-rail approach yields nearly an order-of-magnitude improvement in post-selected infidelity over a single-rail implementation for a given $n$  and parameter set (due to improved infidelities, see Fig.~\ref{fig:infidel_vs_PS}). From the perspective of instead setting a desired infidelity of $1-F<0.2$ and using PS2 estimates, memories of size $N=2^{4}, 2^{8}$ can be queried with single-rail and dual-rail implementations, respectively. 

The price to pay for utilizing post-selection is the rejection of a large fraction of shots. We thus calculate the ``no-flag" probability $P_{\cz}$, the probability that no error is detected during a single $\cz$ gate (of course this is not the no-error probability, as two or more transmon errors can go undetected), see Fig.~\ref{fig:infidel_vs_PS} and Appendix~\ref{app:CZ_fidelity}. For an entire QRAM query to succeed, in the worst case all $\cz$ gates must not flag an error
\begin{align}
P_{\mathrm{no\, flag}} &= \left(P_{\cz}\right)^{N_{\mathrm{gates}}},
\end{align}
with $P_{\mathrm{flag}}=1-P_{\mathrm{no\,flag}}$. The expected time to obtain a successful query is then
\begin{align}
    T_{\rm success} = t_{\cz} N_{\rm ts} / P_{\mathrm{no\, flag}},
\end{align}
where $t_{\cz} N_{\rm ts}$ is the time per query and $1/P_{\mathrm{no\, flag}}$ is the expected number of trials until success. The success rate is then $\Gamma_{\rm success} = 1/T_{\rm success}$.
In the dual-rail case $P_{\mathrm{no\, flag}}$ includes the probability of a photon-loss event for the address or bus qubits during the course of a query. 

The flag probability is higher for dual rail as compared to single rail for a given parameter set, due to the ability to detect cavity photon-loss events [see Fig.~\ref{fig:fidel_and_success}(b)]. This is compensated by the decreased infidelity for dual rail as compared to single rail, see Fig.~\ref{fig:fidel_and_success}(a). For all parameter sets under consideration, the flag probability approaches unity for $n\geq9$. Again considering PS2 estimates, in the single-rail case for $n=4$ we obtain $P_{\mathrm{no\, flag}}=0.71$ (and $\Gamma_{\rm success}=20.9$ kHz), indicating that the majority of queries are expected to succeed. On the other hand, in the dual-rail case for $n=8$, we obtain $P_{\mathrm{no\, flag}} = 1.0\times10^{-6}$ and $\Gamma_{\rm success}=0.013$ Hz. In this case the vast majority of queries are expected to fail, however the fidelity of those that succeed should exceed $F>0.8$. 

It is important to note that our calculated flag probability and success rates are overly pessimistic. In our estimate of $P_{\mathrm{flag}}$, we treat errors at all router locations on equal footing by essentially setting the fidelity to zero in each case. However, a $\cz$ gate failing at the top of the tree is more problematic than one failing at the bottom. This is because queries to all branches through a router that has experienced a gate error are liable to fail \cite{Hann2021}. Thus an error on the top router causes all queries to fail, while an error on a router at the bottom of the tree may only affect queries to that branch (with the necessary caveat that with two-level routers, errors at the bottom of the tree may propagate upward and into other branches as discussed in Sec.~\ref{sec:bb_QRAM} and in Ref.~\cite{Hann2021}). Therefore in practice, $\cz$ gate failures near the bottom of the tree may be tolerated with only a mild reduction in fidelity. Investigation of possible tradeoffs between query fidelity and flag probability is beyond the scope of this work and left for future research.

We also estimate the fidelity of a QRAM query without utilizing post-selection, see the inset of Fig.~\ref{fig:fidel_and_success}(a). High-fidelity queries are only possible for memories of size $2^2=4$, even for our most aggressive coherence time estimates PS3. The reported infidelities are worse than what one might expect from a na\"ive estimate based on the flag probabilities: for instance, in the case of single rail and PS2, the flag probability is only $0.05$ but the non-post-selected infidelity is $1-F=0.26$. The issue is that for small values of $N$, finite-size effects defeat the favorable asymptotic of $\log_{2}^3(N)$. Observe that for $n\leq9$, we have $N_{\mathrm{gates}}<4 N_{\mathrm{ts}}\log_{2}(N)(1+\log_{2}(N))$, despite the linear scaling of $N_{\mathrm{gates}}$ with $N$.

\section{Quantum router via directional photon emission}

\label{sec:GUE}

Recent theoretical~\cite{Guimond2020, Gheeraert2020} and experimental~\cite{Kannan2023} works have investigated {\it giant unidirectional emitters} (GUEs) for use in quantum networks. By coupling two qubits (or cavities) with frequency $\omega$ to a waveguide and spatially separating them by $\lambda/4$, the composite system can be made to emit to the left or right by preparing the system in the state $|\psi_{\rm L}\rangle=(|01\rangle-i|10\rangle)/\sqrt{2}$ or $|\psi_{\rm R}\rangle=(|01\rangle+i|10\rangle)/\sqrt{2}$, respectively~\cite{Gheeraert2020, Guimond2020, Kannan2023}. $\lambda$ here is the wavelength of the emitted photon. We show below that by controlling the direction of emission based on the state of a router (and catching the emitted photon downstream), we implement simultaneous conditional routing operations. This obviates the need we had in the previous model for doing these operations serially.

The simultaneous conditional routing operations are enabled by a pitch-and-catch protocol for quantum state transfer between two GUEs detailed here. 
In a single-rail architecture composed of one GUE (two cavities) at each node, quantum information is routed down the tree in the superposition state
$c_{0}|00\rangle + c_{1}|\psi_{\rm R/L}\rangle$. It is necessary to include the state $|00\rangle$ because we require two states to be routed together that act as a qubit (and the vacuum state is trivially routed in our pitch-and-catch protocol). 
Of course, a photon-loss event from the state $|\psi_{\rm R,L}\rangle$ yields the state $|00\rangle$, amounting to a logical error. This motivates the use of a dual-rail architecture, which utilizes a second pair of GUEs at each send and receive node. The quantum information is then encoded in the superposition state $c_{0}|\psi_{\rm R/L}\rangle|00\rangle + c_{1}|00\rangle|\psi_{\rm R/L}\rangle$. The error state $|00\rangle|00\rangle$ is outside of the codespace and photon-loss events can be detected as in the dual-rail $\cswap$ architecture.

\subsection{Circuit design and state transfer}

\begin{figure*}
    \centering
    \includegraphics[width=2\columnwidth]{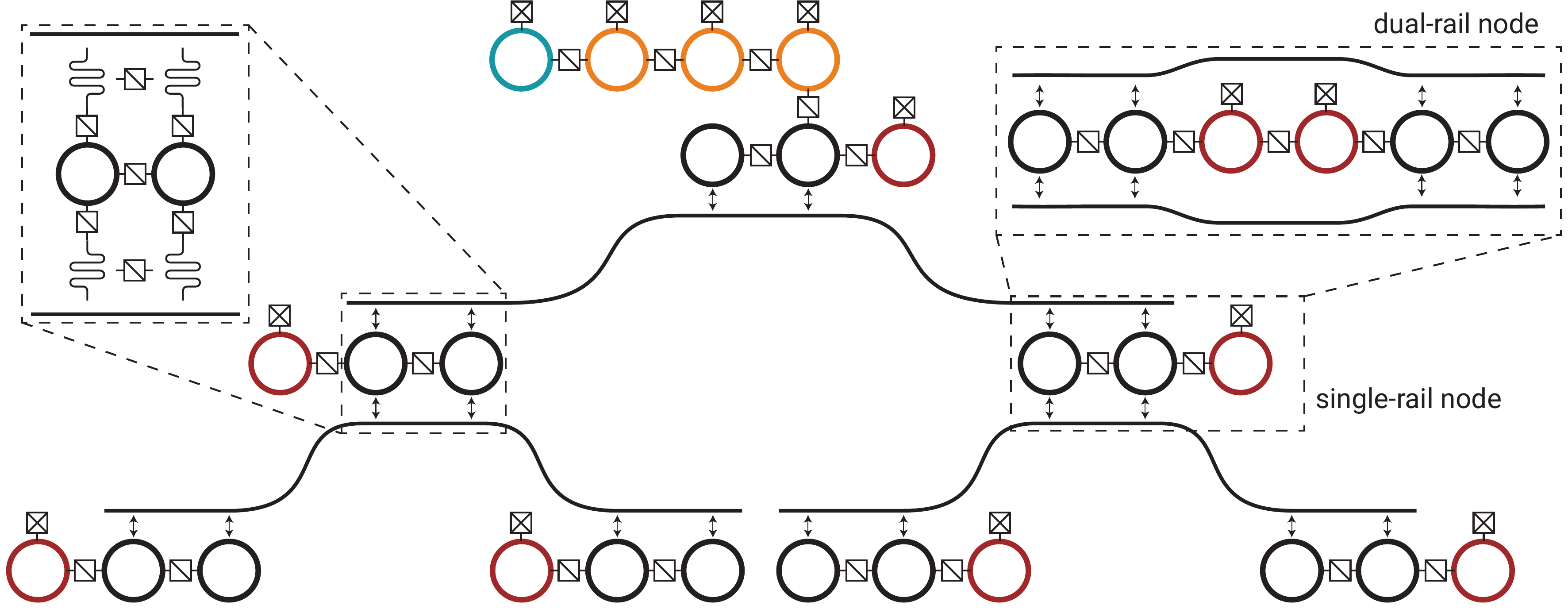}
    \caption{\label{fig:3_bit_GUE} 3-bit QRAM implemented using GUEs. We show explicitly the single-rail architecture, with the right inset detailing the structure of an internal node in the dual-rail case. The left inset indicates the coupling mechanism of the data cavities to the waveguide via transfer resonators.
    }
\end{figure*}

We couple the data cavities indirectly to the waveguide via frequency-converting beamsplitter elements coupled to transfer resonators, see Fig.~\ref{fig:3_bit_GUE}. This coupling architecture has two main advantages. First, frequency-converting beamsplitters allow us to ensure that the emitted photon packets are indistinguishable in frequency, despite the difference in frequency between the data cavities. Previous work~\cite{Gheeraert2020} has shown that the directionality properties of a GUE are highly sensitive to frequency differences between the emitted photons. Second, this architecture allows for an effectively {\it tunable} coupling strength between the data cavities and the waveguide. By turning the coupling strength off, we prevent immediate emission into the waveguide, allowing time for quantum information processing at each node. Once we are prepared to emit a wave packet into the waveguide, we tune the coupling strength to control the shape of the emitted wave packet.
By designing the control pulse to emit a time-symmetric wave packet from a sender GUE, a time-reversed pulse on the receiver GUE absorbs the incident wave packet~\cite{Cirac1997, Korotkov2011, Stannigel2011, Kurpiers2018} . 
We obtain analytical pulse shapes for state transfer in this architecture by adiabatically eliminating the transfer resonators, see Appendix~\ref{app:GUE_state_transfer}. We additionally consider how non-idealities affect the fidelity of state transfer in Appendix~\ref{app:input_output}. For PS2 estimates we predict state-transfer infidelities of $3.3\times10^{-4}$, $4.0\times10^{-6}$ for single rail and dual rail, respectively [see Tab.~\ref{tab:gue_fidel}].  

\begin{table}
\caption{\label{tab:gue_fidel} Coherence times (in addition to those used in Tab.~\ref{tab:params}) used for numerical simulation, infidelities and failure probabilities for state transfer. Transmon coherence times are not relevant here, thus we only consider PS1 and PS2.  }
\begin{ruledtabular}
\begin{tabular}{lccc}
 & parameter set 1  & parameter set 2 \\
\hline 
$T_{1, \mathrm{nr}}^{\,\mathrm{tran}}$ \rule{0pt}{10pt}& 100 $\mu$s  & 200 $\mu$s  \\
$T_{\phi}^{\,\mathrm{tran}}$ & 100 $\mu$s  & 200 $\mu$s \\
$1-F_{\mathrm{st}}$, SR & $1.6\times10^{-3}$  & $3.3\times10^{-4}$ \\
$1-F_{\mathrm{st}}$, DR (p.s.) & $8.3\times10^{-5}$  & $4.0\times10^{-6}$ \\
$1-P_{\mathrm{st}}$, DR & $2.8\times10^{-3}$ & $6.1\times10^{-4}$ \\
\end{tabular}
\end{ruledtabular}
\end{table}

\subsection{GUE-based protocol}

\begin{figure}
    \centering
    \includegraphics[width=\columnwidth]{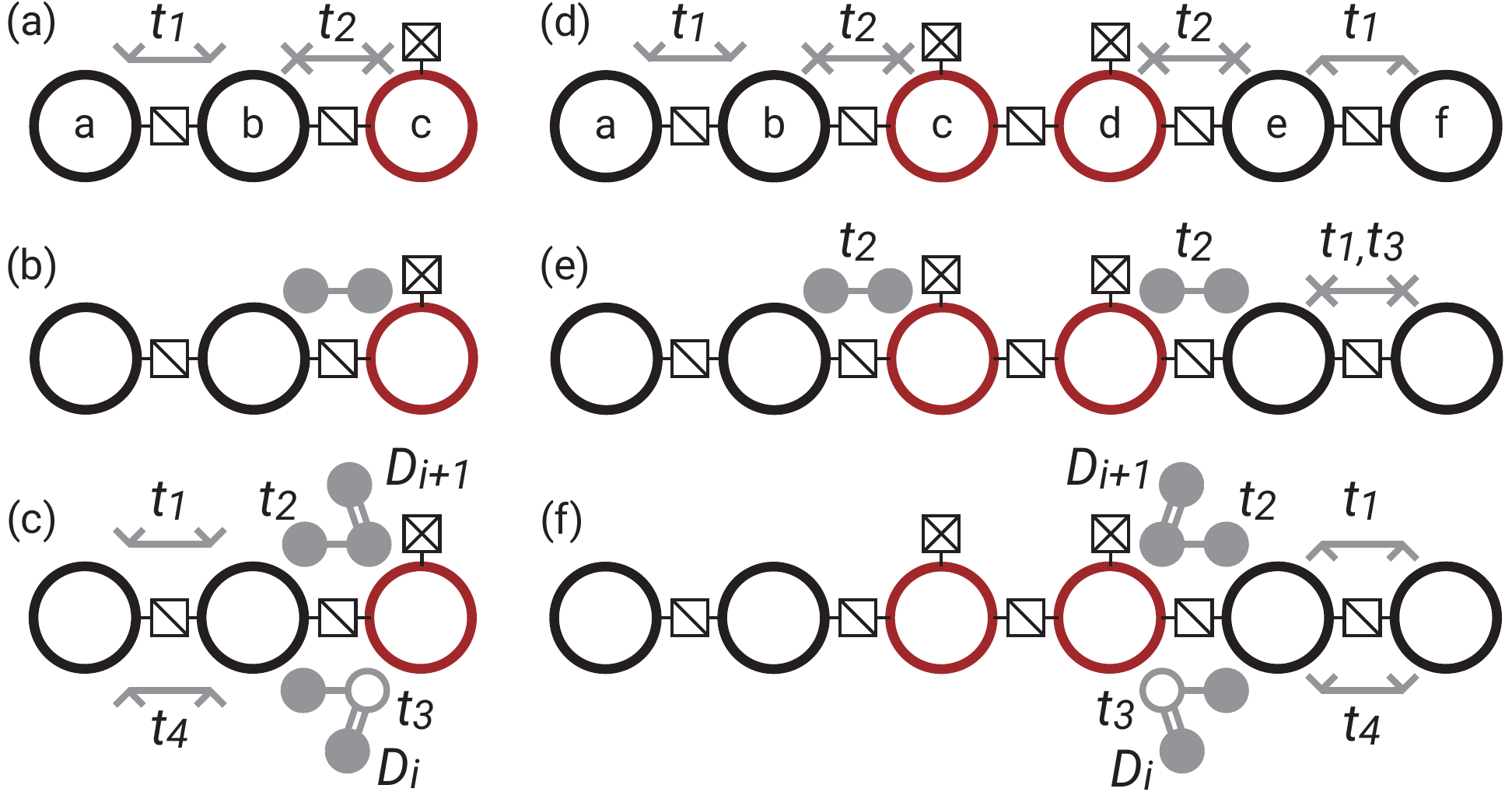}
    \caption{\label{fig:GUE_ops} QRAM operations using GUEs. We have omitted the waveguide and transfer resonators for clarity.
    The three required primitives for bucket-brigade QRAM are (a, d) setting the state of the router, (b, e) controlled routing (the following state-transfer into the waveguide is not shown) and (c, f) copying classical data onto the bus, in the cases of single- and dual-rail, respectively. The gates are labeled $t_{1}, t_{2},\ldots$ to indicate their temporal ordering. 
    Gate schematics and notation utilized here are explained in Fig.~\ref{fig:DR_3_address}(a).}
\end{figure}

We now describe each step of the proposed GUE-based QRAM protocol. As with the $\cswap$ protocol, to perform a bucket-brigade QRAM query we must implement three primitive operations: (i) setting the state of a router, (ii) conditional routing and (iii) encoding classical data in the state of the bus. These operations are shown schematically in Fig.~\ref{fig:GUE_ops} for the single- and dual-rail cases. The operations are similar to those used in the $\cswap$ architecture, with a few differences.

In the case of (i) router-state setting, let us consider a GUE that has just absorbed an incoming wave packet traveling to the right. The state of the GUE+router system is $|\psi\rangle=(\alpha|00\rangle + \beta|\psi_{\rm R}\rangle)|0\rangle$, ordering the states as $a, b, c$ as labeled in Fig.~\ref{fig:GUE_ops}(a). A 50:50 beamsplitter $\mathsf{BS}_{ab}(\pi)$ followed by a single-cavity rotation (done in software) places the address state in cavity $b$ that is nearest the router
\begin{align}
\exp(-i\frac{\pi}{2}\hat{b}^{\dagger}\hat{b})\mathsf{BS}_{ab}(\pi)|\psi\rangle = |0\rangle(\alpha|0\rangle+\beta|1\rangle)|0\rangle.
\end{align}
Finally, a $\mathsf{SWAP}_{bc}$ operation places the address state in the router. These gates (omitting the single-cavity rotation) are shown in Fig.~\ref{fig:GUE_ops}(a). The generalization to the dual-rail case is straightforward and shown in Fig.~\ref{fig:GUE_ops}(d). 

For (ii) conditional routing, we again utilize a phase shift (the logical $\cz$, performed as in Sec.~\ref{sec:DR} for both dual and single rail) to determine the direction in which to send quantum information. Now, the purpose of the $\cz$ operation is to convert e.g. a left-emitting state into a right-emitting state (and vice versa) conditioned on the state of the router \footnote{The router thus serves as a \textquotedblleft switch\textquotedblright  \, that communicates to incoming data either \textquotedblleft keep going\textquotedblright \, or \textquotedblleft change direction.\textquotedblright \, Compiling that information into the actual location encoded by the address qubits can be done easily by tracking phases in software. 
}. In the single-rail case, the operation shown in Fig.~\ref{fig:GUE_ops}(b) yields
\begin{align}
\cz_{bc}[\alpha|00\rangle + \beta|\psi_{\rm R/L}\rangle][ \gamma |0\rangle + \delta |1\rangle ]
= \\ \nonumber \gamma [\alpha|00\rangle + \beta|\psi_{\rm R/L}\rangle]|0\rangle
 + \delta [\alpha|00\rangle + \beta|\psi_{\rm L/R}\rangle]|1\rangle.
\end{align}
Thus, the qubit state is appropriately entangled with the state of the router and emitted right or left accordingly (after the state-transfer protocol is performed). The generalization to the dual-rail case [see Fig.~\ref{fig:GUE_ops}(e)] requires $\cz$ operations on both rails as expected, with additional $\mathsf{SWAP}$ operations and a single-cavity rotation to correct for unwanted phases
\begin{align}
\exp(i\pi\hat{f}^{\dagger}\hat{f})\mathsf{SWAP}_{ef} \cz_{bc}\cz_{de} \mathsf{SWAP}_{ef} \\ \nonumber 
\times[\alpha|\psi_{\rm R/L}\rangle|00\rangle+\beta|00\rangle|\psi_{\rm R/L}\rangle][\gamma|10\rangle+\delta|01\rangle] = \\ \nonumber
\gamma(\alpha|\psi_{\rm L/R}\rangle|00\rangle+\beta|00\rangle|\psi_{\rm L/R}\rangle)|10\rangle \\ \nonumber 
+\delta(\alpha|\psi_{\rm R/L}\rangle|00\rangle+\beta|00\rangle|\psi_{\rm R/L}\rangle)|01\rangle,
\end{align}
ordering the states $a, b, e, f, c, d$ [as labeled in Fig.~\ref{fig:GUE_ops}(d)] such that the router states are on the right.

With respect to (iii) data copying, the protocol is as in Sec.~\ref{sec:DR} with the modification in (i) that a 50:50 beamsplitter and single-cavity rotation are necessary to place the state of the bus in the cavity nearest the router [see Fig.~\ref{fig:GUE_ops}(c, f)]. 

\begin{figure}
    \centering
    \includegraphics[width=\columnwidth]{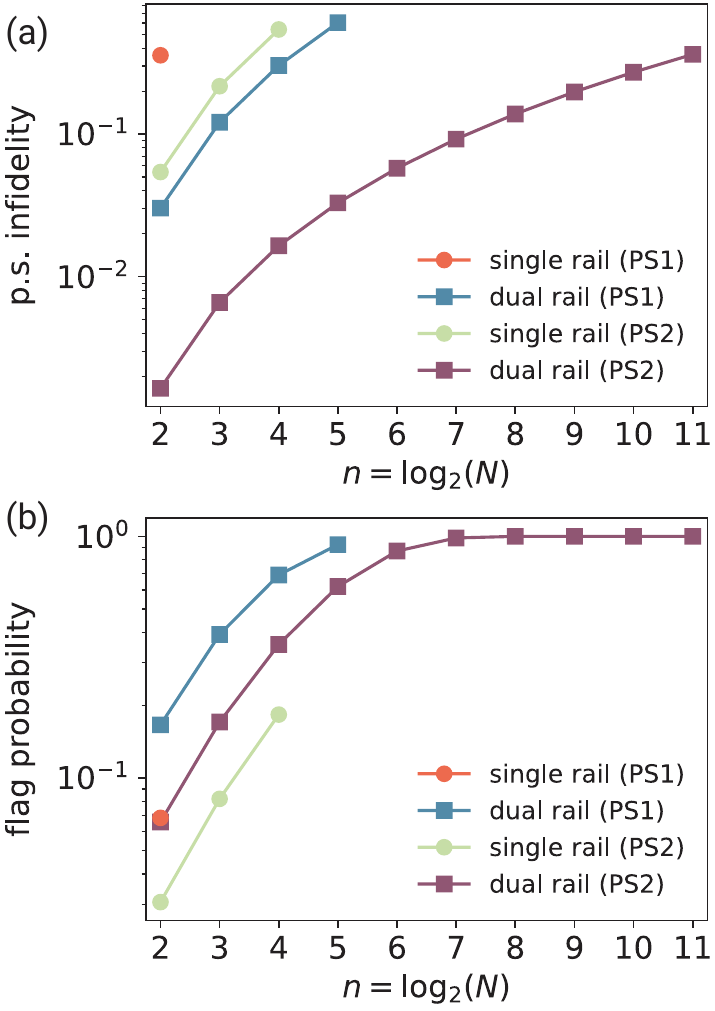}
    \caption{    \label{fig:GUE_fidelity} Post-selected infidelity and failure probability of a QRAM query using GUEs. (a) In the single-rail case, a high-fidelity query is only possible for relatively small $n$ limited by the infidelity of state transfer. In the dual-rail case, we may post-select for photon loss enabling higher-fidelity queries for larger values of $n$. (b) The flag probability in the single-rail case is due to the ability to post-select-away ancilla errors during the $\cz$ operation preceding the state-transfer protocol. }
\end{figure}

\subsection{Resource estimates}

We estimate the hardware cost of the GUE QRAM architecture by counting the number of high-Q resonators, as before. A simple counting argument yields
\begin{align}
N_{\mathrm{cav}}^{\SR} &= 3N+\log_{2}(N)-2, \\ \nonumber 
N_{\mathrm{cav}}^{\DR} &= 2 N_{\mathrm{cav}}^{\SR}.
\end{align}

To estimate the number of gates, we again ignore $\mathsf{SWAP}$ or 50:50 beamsplitter operations between high-Q cavities, focusing on the more costly state transfer and $\cz$ gates. We lump together the (logical) $\cz$ [see Fig.~\ref{fig:GUE_ops}(b, e)] and following state transfer as one ``operation," yielding the total number of gates
\begin{align}
N_{\mathrm{gates}} = 4N - 2 \log_{2}(N) - 4,
\end{align}
valid both for single and dual rail (this number includes the $N$ data copying operations, which do not themselves require a state transfer). The total number of time steps in both cases is
\begin{align}
N_{\mathrm{ts}}=6\log_{2}(N)-6.
\end{align}

\subsection{{Infidelity and error mitigation by post selection}}

The error probability associated with each time step is $\epsilon = 1 - F_{g}(\cz)F_{\mathrm{st}}$, grouping the router-controlled operations and following state transfer together, and defining the fidelity of state transfer $F_{\rm st}$. This expression for the error probability assumes that errors are uncorrelated between the two operations. This may not be exactly true in practice, however it is likely a good leading-order approximation and significantly reduces the computational cost of calculating the error probability. We additionally define the no-flag probability
\begin{align}
P_{\rm no\,flag} = (P_{\cz}P_{\rm st})^{N_{\rm gates}},
\end{align}
where $P_{\rm st}$ is the success probability for the state-transfer procedure. $P_{\rm st}$ is unity in the single-rail case, however it takes nontrivial values in the dual-rail case because photon loss is detectable. Details on the calculation of $F_{\rm st}$ and $P_{\rm st}$ are provided in Appendix~\ref{app:GUE_state_transfer}. To calculate the overall infidelity of a QRAM query we again use Eq.~\eqref{eq:fidel_two_level_router}, see Fig.~\ref{fig:GUE_fidelity}. 

For PS2 estimates, in the single-rail case only relatively small-scale devices are predicted to yield high fidelities. The issue is that for the state-transfer protocol considered here, we predict state-transfer infidelities on the order of $10^{-4}$. This contribution to the overall infidelity overwhelms that due to the $\cz$ operation, which has a predicted post-selected infidelity on the order of $10^{-5}$ for PS2 estimates, see Fig.~\ref{fig:infidel_vs_PS}. 

The situation is quite different in the dual-rail case, with post-selected query fidelities of $F\geq 0.9$ possible for $n\leq6$ and using PS2 estimates. In the case of $n=8$, we obtain $F>0.8$, $P_{\rm no\,flag}=2\times10^{-4}$ and $\Gamma_{\rm success}=2.2$ Hz. The difference now is that photon loss during the state-transfer procedure is detectable, as the vacuum state is no longer a logical state. The price to be paid is the increased flag probability of dual rail with respect to single rail (for a given parameter set), see Fig.~\ref{fig:GUE_fidelity}(b). 

\section{Discussion}

\label{sec:disc}

We have explored two approaches towards realizing QRAM using superconducting circuits. The first relies on the direct construction of $\czeroswap$ and $\coneswap$ gates between three superconducting 3D cavities. 
The only gate primitive required (aside from cavity $\mathsf{SWAP}$s and beamsplitters) is the $\mathsf{ZZ}(\pi/2)$ gate, which underlies both the conditional routing and data copying operations. Such a limited required gate set could significantly alleviate control and calibration overhead in a potential experimental realization. Additionally, this architecture has the advantage of requiring no new hardware beyond what has already been experimentally realized in e.g. Refs.~\cite{Lu2023, Chapman2022, Chou2023}. The experimentally feasible architecture and small gate set stand in contrast to previous QRAM proposals, which either still require the experimental implementation of elementary hardware components \cite{Giovannetti2008architectures, Hong2012, Hann2019, vonLupke2023}, or utilizes components which have not yet been demonstrated together \cite{Chen2021}.

The second architecture introduces GUEs, which allow for the simultaneous conditional routing of quantum information down the tree in both directions (whereas in the $\cswap$ architecture, the routing operations must be performed serially). The coupling of 3D cavities to a waveguide in the manner described in this work has yet to be realized experimentally. However, the tunable coupling of 2D transmons to a waveguide has been experimentally demonstrated~\cite{Kannan2023}. Many of our results carry over immediately to this case, most notably the state-transfer protocol which is carried out in the single-excitation manifold. This platform could then implement the QRAM protocol described in this work, at the cost of using relatively low-coherence transmons as routers. 

{It is worth emphasizing that in both architectures, the classical data is not stored in any quantum object before it is copied into the state of the bus. Instead, the classical data determines which gates are performed during the QRAM query. This simplification results in quantum hardware savings, and ensures that we need not be concerned with decoherence in ``memory cell qubits" \cite{Kim2023}, which do not exist in our proposal. }

{
On the one hand, the GUE architecture generally enjoys a success-rate advantage over the $\cswap$ architecture. This is due to the fact that the lossy ancilla is only meaningfully excited during the single $\cz$ gate preceding the state-transfer protocol, and not during the state transfer itself. This is to be contrasted with the $\cswap$ case, where two $\cz$ gates are required at each node to perform the conditional routing. On the other hand, the error per timestep is higher in the single-rail-GUE case than for single-rail-$\cswap$ due to the limited fidelity of the state-transfer protocol as compared to the postselected $\cz$ gates (the postselected state-transfer fidelity is improved in the dual-rail case). Thus in attempting to build a QRAM using one of the proposed architectures, the experimenter must carefully consider the tradeoffs between hardware complexity, fidelity, success rate, etc.
}


One could additionally envision merging the two proposed architectures. Laying out the QRAM in a tree-like structure, nodes closest to the root are naturally further spatially separated from one another. This motivates perhaps connecting these nodes via the GUE architecture, before reverting to beamsplitter connectivity for the nodes nearer the bottom. Of course, due to the use of 3D cavities with centimeter-scale footprints, scaling up our QRAM architectures will be challenging due to the limited space available in dilution refrigerators. However, we expect small-scale devices to be realizable in the near term \cite{Chou2023}. These proof-of-principle experiments will motivate exploring strategies for scaling up if they prove successful.

Recent work on {\it virtual QRAM}~\cite{Xu2023} has explored implementing an $N$-bit QRAM on hardware nominally supporting only an $M$-bit query, with $M<N$. This technique comes at the cost of increased noise sensitivity, though Xu {\it et al.} show the robustness of virtul QRAM to $\textit{Z}$-biased noise~\cite{Xu2023}. For the 3D cavities considered here, the main noise channel is amplitude damping via photon loss~\cite{Rosenblum2018, Sivak2023}. Whether the virtual-QRAM technique can be made robust to these errors is left for future work. 

It is worthwhile to place our work in the context of modern quantum algorithms that utilize QRAM. 
Two examples of recent interest include (i) the quantum simulation of jellium with at least 54 electrons studied in Ref.~\cite{Babbush2018} and (ii) the factoring of RSA-2048 with a modern version of Shor's algorithm analyzed in Ref.~\cite{Gidney2021}. In (i) a QRAM of size $N\geq162$ (at least $n=8$ address qubits) is required for loading Hamiltonian coefficients.  This device is queried on the order of $10^4$ times, with a bus register of about 13 qubits~\cite{Babbush2018}. In (ii), a QRAM of size $\sim2^{10}$ is utilized for classical precomputation of modular exponentials. This device is queried on the order of $10^5$ times and requires a bus register of 2048 qubits (the number of bits of the integer to be factored). 
These results suggest that in terms of the number of address qubits, QRAMs of the sizes considered in this work may be relevant for quantum algorithms showing quantum advantage. However, query infidelities must decrease at least to the $10^{-4}-10^{-5}$ level to support $\geq10^4-10^5$ QRAM queries. Additionally, in our architecture each qubit in the bus register must be sent into the QRAM one at a time. This lengthens the query time and decreases the fidelity of a query. It is thus worthwhile in future work to consider noise-resilient QRAM architectures that natively support large bus registers.

{Further future work could explore larger-scale numerical simulations \cite{Bugalho2023} that go beyond a calculation of the error per timestep $\epsilon$ and use of the query infidelity formula~\eqref{eq:fidel_two_level_router}. Such simulations may be possible due to the limited entanglement present in the QRAM device, allowing for the use of e.g. tensor-network methods. Such simulations would allow for the investigation of effects such as differing decoherence rates and measurement infidelities among different routers.
}

\section{Acknowledgments}

We thank Stijn J. de Graaf for a critical reading of the manuscript. We thank Stijn J. de Graaf, Yongshan Ding, Bharath Kannan, Shifan Xu, and Sophia Xue for helpful discussions. We acknowledge use of the Grace cluster at the Yale Center for Research Computing. This material is based upon work supported by the Air Force Office of Scientific Research under award number FA9550-21-1-0209. The U.S. Government is authorized to reproduce and distribute reprints for Governmental purposes notwithstanding any copyright notation thereon. The views and conclusions contained herein are those of the authors and should not be interpreted as necessarily representing the official policies or endorsements, either expressed or implied, of the Air Force Office of Scientific Research or the U.S. Government.

S.P. and S.M.G. receive consulting fees from Quantum Circuits, Inc. S.M.G. is an equity holder in Quantum Circuits, Inc.

\appendix

\section{Quantum circuit for 3-bit QRAM}
\label{app:qcircuit}

\begin{figure*}
    \centering
    \includegraphics{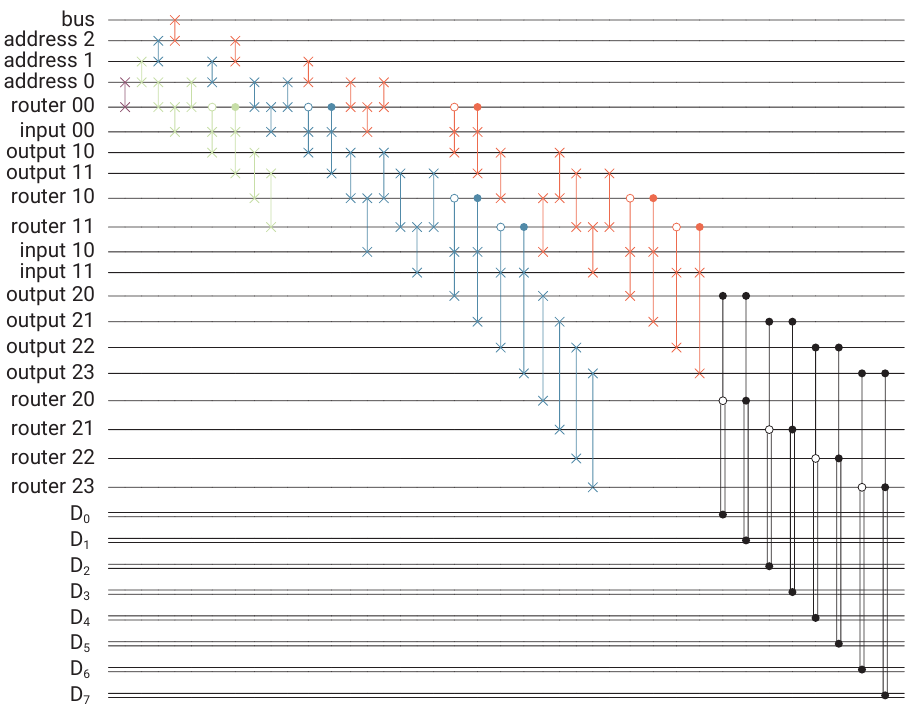}
    \caption{\label{fig:qcircuit_3_bit} Quantum circuit for a query on the 3-bit QRAM of Fig.~\ref{fig:DR_3_address}(e). The gates required to route in address 0, address 1,address 2, and the bus are colored as purple, green, blue, and orange, respectively. The gates in black copy classical data into the bus. Referring to the routing gates (purple, green, blue and orange gates) as $U_{\rm route}$ and the data copy gates (black) as $U_{\rm data}$, a full QRAM query is given by the unitary $U_{\rm route}^{\dagger}U_{\rm data}U_{\rm route}$. The gates in $U_{\rm route}^{\dagger}$ which disentangle the bus from the routers are not shown due to space limitations. This quantum circuit was typeset using the QCircuit package~\cite{qcircuit}. }
\end{figure*}

The full quantum circuit associated with a query on the 3-bit QRAM of Fig.~\ref{fig:DR_3_address}(e) is shown in Fig.~\ref{fig:qcircuit_3_bit}. All of the gates are local and color coded for clarity. We observe the benefits of address pipelining~\cite{Xu2023, Jaques2023}, where the third address qubit and the bus qubit undergo simultaneous $\cswap$ operations in different layers of the QRAM device.

\section{CSWAP operation}
\label{app:DR_CSWAP}

For the computational basis states $|n_{a}00\rangle, |n_{a}01\rangle, |n_{a}10\rangle, n_{a}=0,1$ the operation  $\mathsf{BS}_{bc}^{\dagger}(\pi/2)\cz_{ab}\mathsf{BS}_{bc}(\pi/2)$ appropriately functions as a $\textsf{CSWAP}$ operation. To see this, note first that when the target modes are unoccupied, the gate operation is trivial $\mathsf{BS}_{bc}^{\dagger}(\pi/2)\cz_{ab}\mathsf{BS}_{bc}(\pi/2)|n_{a}00\rangle=|n_{a}00\rangle$. Now if the $b$ mode is occupied, we obtain
\begin{align}
&\mathsf{BS}_{bc}^{\dagger}(\pi/2)\cz_{ab}\mathsf{BS}_{bc}(\pi/2)|n_{a}10\rangle = \\ \nonumber 
&\frac{1}{2}[1+(-1)^{n_{a}}]|n_{a}10\rangle+\frac{1}{2}[1-(-1)^{n_{a}}]|n_{a}01\rangle.
\end{align}
Thus if $n_{a}=0$, no $\mathsf{SWAP}$ is performed, while if $n_{a}=1$, the photon is swapped to mode $c$. Similarly if mode $c$ is initially occupied and mode $b$ is unoccupied, we obtain
\begin{align}
&\mathsf{BS}_{bc}^{\dagger}(\pi/2)\cz_{ab}\mathsf{BS}_{bc}(\pi/2)|n_{a}01\rangle= \\ \nonumber 
&\frac{1}{2}[1-(-1)^{n_{a}}]|n_{a}10\rangle+\frac{1}{2}[1+(-1)^{n_{a}}]|n_{a}01\rangle.
\end{align}
We have thus obtained an effective $\textsf{CSWAP}$ operation considering only the relevant states.

In the case of the initial state $|n_{a}11\rangle$, $n_{a}=0,1$ the 50:50 beamsplitter yields
\begin{align}
\mathsf{BS}_{bc}(\pi/2)|n_{a}11\rangle = \frac{1}{\sqrt{2}}(|n_{a}02\rangle-|n_{a}20\rangle),
\end{align}
where the photons bunching in one or the other mode is known as the Hong-Ou-Mandel effect.
Now, the state $|n_{a}20\rangle$ outside of the computational manifold incurs a phase shift as a result of the $\cz$ operation \big(see Eq.~\eqref{eq:CZ}, contrast with the case of a pure Kerr interaction $\exp[i\pi \hat a^{\dagger}\hat a\hat b^{\dagger}\hat b]$\big)
\begin{align}
\cz_{ab} \frac{1}{\sqrt{2}}(|n_{a}02\rangle-|n_{a}20\rangle) = \frac{1}{\sqrt{2}}(|n_{a}02\rangle+|n_{a}20\rangle).
\end{align}
The resulting state is a dark state of the inverse beamsplitter
\begin{align}
\mathsf{BS}_{bc}^{\dagger}(\pi/2)\frac{1}{\sqrt{2}}(|n_{a}02\rangle+|n_{a}20\rangle) = \frac{1}{\sqrt{2}}(|n_{a}02\rangle+|n_{a}20\rangle),
\end{align}
and thus population in the state $|n_{a}11\rangle$ is transferred out of the computational manifold at the completion of the gate $\mathsf{BS}_{bc}^{\dagger}(\pi/2)\cz_{ab}\mathsf{BS}_{bc}(\pi/2)$. In the case of the ideal Kerr interaction there is no such phase shift on the $|n_{a}20\rangle$ state and the inverse beamsplitter undoes the Hong-Ou-Mandel effect. We emphasize again that we do not expect both target modes to be occupied in the course of a QRAM query, as the system is initialized in the vacuum state. 

\section{Fidelity of the $\cz$ operations}
\label{app:CZ_fidelity}

Based on the ability to perform first-order error detection, 
we want to calculate the post-selected fidelity of the $\cz$ gate. To proceed, we define measurement operators $\mathsf{M}_{\checkmark}, \mathsf{M}_{\times}$. A measurement result associated with $\mathsf{M}_{\checkmark}$ indicates no error was detected, while $\mathsf{M}_{\times}$ indicates an error was detected. If before measurement the system was in the state $\rho$, the post-measurement state $\rho'$ conditioned on detecting no errors is
\begin{align}
\label{eq:rho_measure}
\rho' = \mathsf{M}_{\checkmark}\rho \mathsf{M}_{\checkmark}^{\dagger}/P_{\checkmark}(\rho),
\end{align}
where $P_{\checkmark}(\rho)=\Tr(\rho \mathsf{M}_{\checkmark}^{\dagger}\mathsf{M}_{\checkmark})$ is the probability of no errors. As discussed in Refs.~\cite{Teoh2022, Tsunoda2022}, measurement of the ancilla transmon after the $\cz$ operation provides first-order detection of errors in the ancilla (in the dual-rail case, we measure both transmons associated with the router rails). Measuring the transmon in $|g\rangle$ indicates no errors were detected, while measurement in $|e\rangle, |f\rangle$ indicate a bit-flip or a phase-flip error, respectively. (The gate of course may still fail, if e.g., two ancilla errors occur during the gate.)
{In experimental reality, some jumps may occur that do not get flagged. That is to say, we may misidentify the $|e\rangle,|f\rangle$ states as $|g\rangle$. We model this by taking as our no-error measurement operator 
\begin{align}
\mathsf{M}_{g}=\sqrt{\eta_{gg}}|g\rangle\langle g|+\sqrt{\eta_{ge}}|e\rangle\langle e|+\sqrt{\eta_{gf}}|f\rangle\langle f|,    
\end{align} 
with measurement coefficients $\eta_{gg}=1-10^{-4}, \eta_{ge}=0.01, \eta_{gf}=\eta_{ge}^2$~\cite{Tsunoda2022}. (The opposite problem of misidentifying $|g\rangle$ as $|e\rangle, |f\rangle$ only serves to decrease the success rate). We can already see that such detrimental effects only enter at second order: first an error must occur, then it must be misidentified (rare for cQED experiments with high-fidelity readout such as e.g. Ref.~\cite{Sivak2023}). 
}

In the dual-rail case we can also detect photon-loss events in the cavities by performing a parity check~\cite{Teoh2022}.
It is important to note that this parity check can be done only at the end of the QRAM query, once the addresses and bus have been routed in and back out of the tree \footnote{Note that the QRAM query times $t_{n}=t_{\cz} N_{\rm ts}$ are shorter than $T_{1}^{c}$ for the parameters considered in this paper. Thus a parity check performed only at the end of the circuit is sufficient, and we do not expect to be limited by uncaught cavity decay/heating events~\cite{Levine2023}}. Otherwise, we reveal ``which path" information which destroys the superposition state.
Thus in the following we do not explicitly simulate a parity check at the completion of the $\cz$ operation, which would take additional time $\sim\pi/\chi$~\cite{Teoh2022, Tsunoda2022} and unrealistically increase our error probability per time step due to additional uncaught transmon errors. Instead, we perform the ancilla measurement(s) as described above, then project onto the dual-rail basis to obtain an estimate of the fidelity boost due to utilizing dual-rail qubits. The overall no-error measurement operator is 
\begin{align}
\mathsf{M}_{\checkmark} = \mathsf{M}_{\DR}  \mathsf{M}_{g},
\end{align}
where 
\begin{align}
\mathsf{M}_{\DR}=\sum_{i,j=0}^{1}|i_{\mathrm{L}}\rangle|j_{\mathrm{L}}\rangle\langle i_{\mathrm{L}}|\langle j_{\mathrm{L}}|,
\end{align}
written in the logical dual-rail basis. This expression for $\mathsf{M}_{\DR}$ ignores measurement errors in projecting onto the dual-rail basis, however such effects are sub-leading order compared to ancilla measurement errors and thus can be safely ignored. 
We have additionally neglected the effects of no-jump backaction, which arises if the decay rates are not identical between the two cavities comprising a dual rail \cite{Tsunoda2022}. Essentially, superposition states become biased toward occupation of the longer lived cavity. This effect can be mitigated by periodically performing $\swap$s between the two rails of a dual rail, effectively echoing out the no-jump backaction.

The normalization by $P_{\checkmark}(\rho)$ in Eq.~\eqref{eq:rho_measure} ensures that the density matrix $\rho'$ has unit trace. This nonlinear mapping complicates the application of standard fidelity metrics, which assume a linear quantum channel~\cite{Nielsen2002, Dankert2005} that can be decomposed into a set of state-independent Kraus operators~\cite{Pedersen2007}.
To proceed, we instead view this process as a linear map that yields a subnormalized state. Thus, we can apply formulas that depend on the process being a quantum channel, before correcting for the subnormalization. We utilize Nielsen's formula for the entanglement fidelity associated with a gate $\mathsf{U}$~\cite{Nielsen2002}
\begin{align}
\widetilde{F}_{e}(\mathsf{U}) &= \frac{\sum_{jk}\alpha_{jk}\Tr(\mathsf{U} \mathsf{U}_{j}^{\dagger}\mathsf{U}^{\dagger}[\mathsf{M}_{\checkmark}\mathcal{E}_{\mathsf{U}}\{\rho_{k}\}\mathsf{M}_{\checkmark}^{\dagger}])}{d^3},
\end{align}
where $\mathcal{E}_{\mathsf{U}}(\rho)$ is the quantum channel applied to the density matrix $\rho$ before the error-detecting measurements, and $\mathcal{E}_{\mathsf{U}}(\rho) = \mathsf{U} \rho \mathsf{U}^{\dagger}$ when the map is unitary. $d$ is the dimension of the Hilbert space (here $d=4$), the $\mathsf{U}_{j}$ are an orthonormal operator basis on the $d$-dimensional space, and the $\rho_{k}$ are pure state density matrices consisting of the computational basis states $|0\rangle, \ldots, |d-1\rangle$ as well as their superpositions $(|j\rangle\pm|k\rangle)/\sqrt{2}$ and $(|j\rangle\pm i|k\rangle)/\sqrt{2}$ for $j\neq k$. (In the case $d=4$, there are 4 computational basis states and 24 superposition states.) The coefficients $\alpha_{jk}$ are defined by the decomposition $\mathsf{U}_{j}=\sum_{k}\alpha_{jk}\rho_{k}$. It is convenient to use the basis $\sqrt{d}|j\rangle\langle k |$, as the decomposition in terms of density matrices is simple $|j\rangle\langle k | = |+\rangle\langle+|+i|-\rangle\langle-|-[|j\rangle\langle j|+|k\rangle\langle k|](1+i)/2$, where $|+\rangle=(|j\rangle+|k\rangle)/\sqrt{2}$ and $|-\rangle=(|j\rangle+i|k\rangle)/\sqrt{2}$~\cite{Mohseni2008} (note the typo in the formula in Ref.~\cite{Mohseni2008}, fixed here).

The quantity $\widetilde{F}_{e}(\mathsf{U})$ encodes the product of the average entanglement fidelity and the success probability, as opposed to the average entanglement fidelity alone. We thus divide by the average success probability $P_{\mathsf{U}}=\sum_{k}\Tr(\mathcal{E}_{\mathsf{U}}(\rho_{k}))/[2d^2-d]$ to obtain the average post-selected entanglement fidelity $F_{e}(\mathsf{U})=\widetilde{F}_{e}(\mathsf{U})/P_{\mathsf{U}}$, summing over the $2d^2-d$ states utilized in the decomposition of the operator basis.
The average post-selected gate fidelity $F_{g}(\mathsf{U})$ is then given by the standard formula~\cite{Nielsen2002, Horodecki1999}
\begin{align}
F_{g}(\mathsf{U}) = \frac{d F_{e}(\mathsf{U}) + 1}{d+1}.
\end{align}

\begin{figure}
    \centering
    \includegraphics[width=\columnwidth]{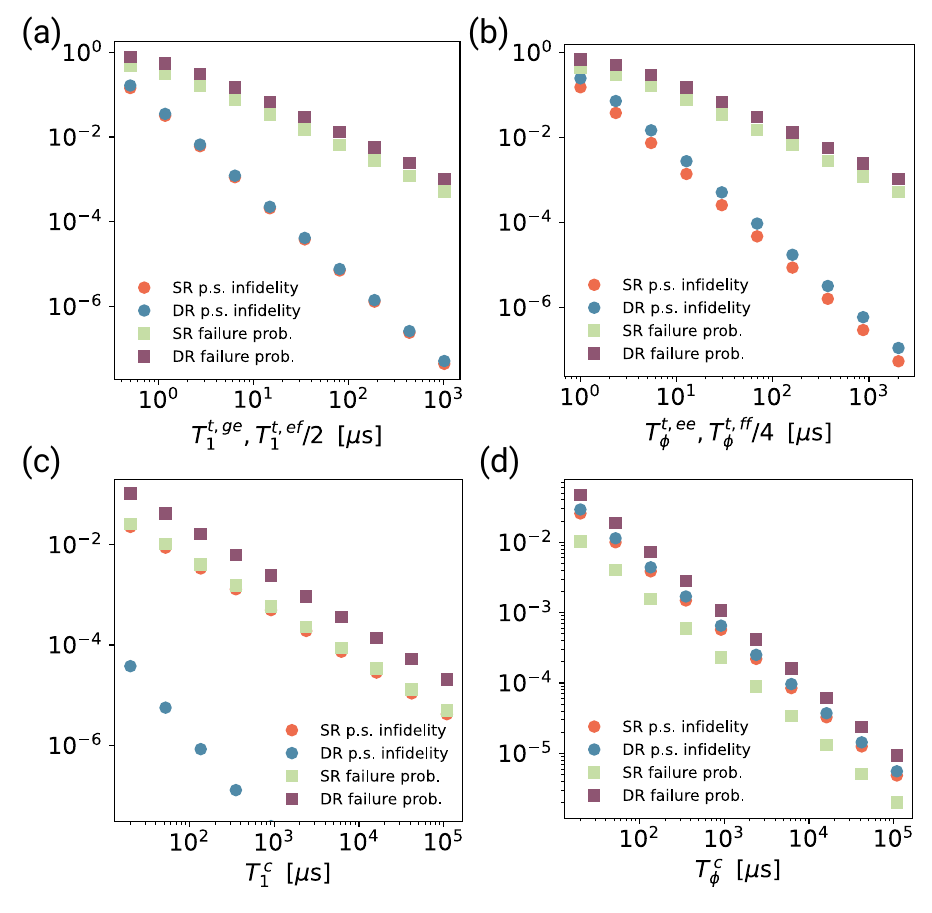}
    \caption{\label{fig:bosonic_fidel_vs_coherence}
    Post-selected gate infidelity and failure probability of the $\cz$ operation. We sweep over transmon (a) relaxation and (b) dephasing times as well as cavity (c) relaxation and (d) dephasing. To observe the scaling of the infidelity and failure probability with each parameter, we set all decay rates to zero that are not being swept. In the single-rail case, post-selection allows for the detection of first-order ancilla decay and dephasing errors (a-b). Thus the post-selected infidelities scale approximately quadratically with the coherence times. In the dual-rail case, we may now additionally post-select on first-order cavity-decay errors (c). 
}
\end{figure}

In the single-rail case we simulate the $\cz$ operation as described in Sec.~\ref{sec:DR} and Ref.~\cite{Tsunoda2022}, utilizing the QuTiP~\cite{qutip1, qutip2} Lindblad master equation solver. We observe that the post-selected infidelity scales with $(T_{1}^{t, ge})^{-2}, (T_{\phi}^{t, ee})^{-2}$ due to the ability to detect first-order transmon errors, see Fig.~\ref{fig:bosonic_fidel_vs_coherence}(a)-(b). Of course, the failure probability scales as $(T_{1}^{t, ge})^{-1}, (T_{\phi}^{t, ee})^{-1}$. Additionally, in the single-rail case there is no ability to detect first-order photon loss errors in the cavities (dephasing events in the cavity cannot be detected in either case). Thus the post-selected infidelities and failure probabilities both scale as $(T_{1}^{c})^{-1}, (T_{1}^{\phi})^{-1}$.

In the dual-rail case, we make the simplification that the two halves of the dual rails are identical. This allows us to reuse results from the single-rail case, and appropriately tensor together single-rail states to obtain dual-rail states. In this way we simulate the logical $\mathsf{CZ}$ operation, which consists of two parallel physical $\cz$ gates as in the logical $\textsf{CSWAP}$s schematically represented in Fig.~\ref{fig:DR_3_address}(c). Now with the ability to detect first-order decay events in the cavities, we obtain a post-selected infidelity that scales with $(T_{1}^{c})^{-2}$ [see Fig.~\ref{fig:bosonic_fidel_vs_coherence}(c)].
Dephasing events in the cavity are still undetectable, thus the post-selected infidelity still scales as $(T_{\phi}^{c})^{-1}$ [see Fig.~\ref{fig:bosonic_fidel_vs_coherence}(d)].

\section{GUE state transfer}
\label{app:GUE_state_transfer}

The state transfer problem involves the sender GUE emitting in both directions simultaneously in superposition. For simplicity we restrict ourselves to the problem of state transfer between only two GUEs;  we show below that in the ideal case of parameter symmetry, the state-transfer problems for both directions decouple. (When performing numerics and including decoherence processes, we analyze the full state-transfer problem of six data cavities and six transfer resonators.)

\begin{figure}
    \centering
    \includegraphics[width=\columnwidth]{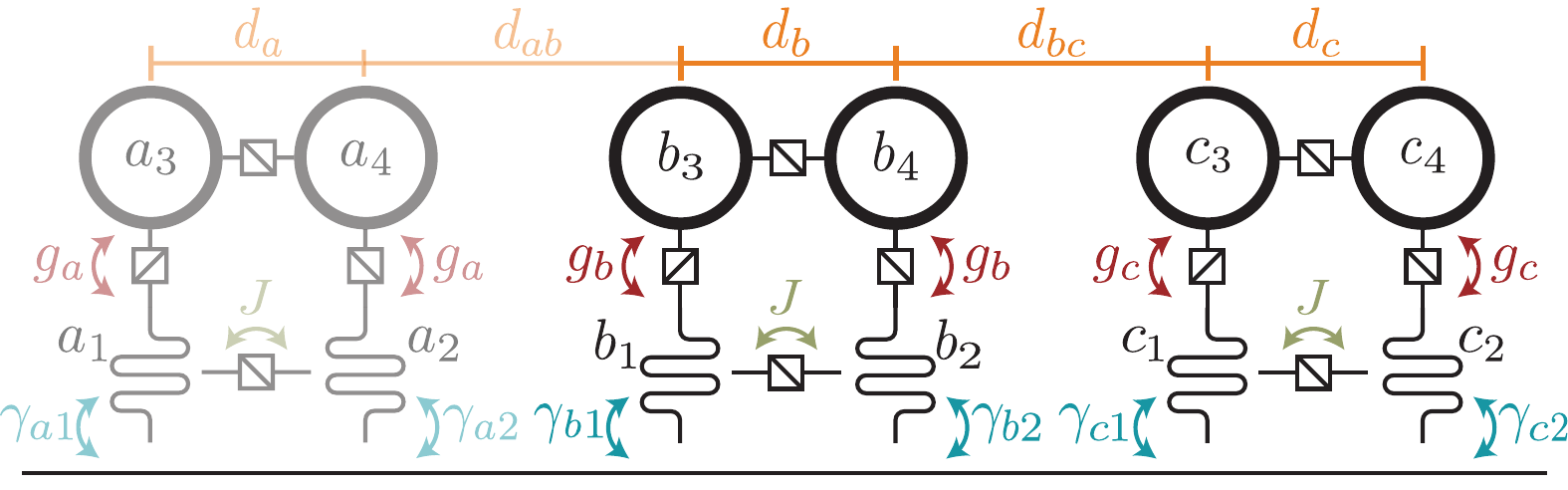}
    \caption{\label{fig:fig10} Schematic of multiple GUEs coupled to a common waveguide. The physics of the state-transfer protocol can be understood by considering only two pairs of GUEs coupled to the waveguide, e.g. GUEs $b$ and $c$.}
\end{figure}
The bare Hamiltonian of the waveguide and the GUEs is
\begin{align}
\label{eq:H0_GUE}
\hat{H}_{0} = &\int_{0}^{\infty}\mathrm{d}\omega\, \omega [\hat{a}_{\rm R}^{\dagger}(\omega)\hat{a}_{\rm R}(\omega)+\hat{a}_{\mathrm{L}}^{\dagger}(\omega)\hat{a}_{\mathrm{L}}(\omega)] \\ \nonumber 
 &+ \sum_{\substack{\mu=b,c\\j=1,2,3,4}}\omega_{\mu j}\hat{\mu}_{j}^{\dagger}\hat{\mu}_{j},
\end{align}
where $\hat{a}_{\mathrm{L}}(\omega), \hat{a}_{\rm R}(\omega)$ are the annihilation operators of left and right propagating modes at of the waveguide at frequency $\omega$, respectively, $\hat{\mu}_{1}, \hat{\mu}_{2}$ are the annihilation operators of the left and right transfer resonators, respectively, and $\hat{\mu}_{3}, \hat{\mu}_{4}$ are the annihilation operators of the left and right data cavities, respectively [see Fig.~\ref{fig:fig10}].
It is important to note that the waveguide is terminated on each end with a $50 \;\Omega$ impedance, such that the waveguide supports a continuum of modes as in Eq.~\eqref{eq:H0_GUE}.

Including now the tunable beamsplitter interactions and moving into the interaction picture with respect to $\hat{H}_{0}$, the intra-GUE Hamiltonian is [making the rotating-wave approximation (RWA)]
\begin{align}
\label{eq:Hsys}
\hat{H}_{\mathrm{sys}}(t) = \sum_{\mu=b, c}[g_{\mu}(t)(\hat{\mu}_{1}^{\dagger}\hat{\mu}_{3} + \hat{\mu}_{2}^{\dagger}\hat{\mu}_{4}) + J \hat{\mu}_{1}^{\dagger}\hat{\mu}_{2} +  \mathrm{H. c.}].
\end{align}
The coefficients $g_{\mu}(t)$ encode the time-dependent drive envelope of the beamsplitter interaction between the data cavities and the transfer resonators [see Fig.~\ref{fig:fig10}]. As noted in Refs.~\cite{Guimond2020,Gheeraert2020,Kannan2023}, the static interaction between the transfer resonators is necessary to cancel an effective unwanted exchange interaction mediated by the waveguide. Of course, a beamsplitter interaction between the data cavities [omitted in Eq.~\eqref{eq:Hsys}, as it is not relevant to the present discussion] is also necessary, e.g. for state preparation.
The interaction Hamiltonian between the GUEs and the waveguide is~\cite{Guimond2020}
\begin{align}
\nonumber
\hat{H}_{\rm I} &=\frac{1}{\sqrt{2\pi}}\int_{0}^{\infty}d\omega e^{i(\omega-\omega_{0})t} \{\hat{a}_{\mathrm{L}}^{\dagger}(\omega)[\hat{L}_{\mathrm{L}}^{(b)}(\omega)+\hat{L}_{\mathrm{L}}^{(c)}(\omega)] \\  
\label{eq:H_I}
&\quad +\hat{a}_{\rm R}^{\dagger}(\omega)[\hat{L}_{\rm R}^{(b)}(\omega)+\hat{L}_{\rm R}^{(c)}(\omega)]\}+\mathrm{H.c.},
\end{align}
where
\begin{align}
\label{eq:L_defs}
\hat{L}_{\rm R}^{(b)}(\omega) &= \sqrt{\gamma_{b1}}\hat{b}_{1}+e^{-i\omega d_{b}/v}\sqrt{\gamma_{b2}}\hat{b}_{2}, \\ \nonumber 
\hat{L}_{\mathrm{L}}^{(b)}(\omega) &= \sqrt{\gamma_{b1}}\hat{b}_{1}+e^{i\omega d_{b}/v}\sqrt{\gamma_{b2}}\hat{b}_{2}, \\ \nonumber 
\hat{L}_{\rm R}^{(c)}(\omega) &= e^{-i\omega (d_{b}+d_{bc})/v}(\sqrt{\gamma_{c1}}\hat{c}_{1}+e^{-i\omega d_{c}/v}\sqrt{\gamma_{c2}}\hat{c}_{2}), \\ \nonumber  
\hat{L}_{\mathrm{L}}^{(c)}(\omega) &= e^{i\omega (d_{b}+d_{bc})/v}(\sqrt{\gamma_{c1}}\hat{c}_{1}+e^{i\omega d_{c}/v}\sqrt{\gamma_{c2}}\hat{c}_{2}).
\end{align}
We define the decay rate $\gamma_{\mu j}$ of transfer resonator $\mu_{j}, \,j=1,2$ into the waveguide, see Fig.~\ref{fig:fig10}. The separation between transfer resonators in GUEs $b$ and $c$ are defined as $d_{b}$ and $d_{c}$, respectively, while $d_{bc}$ denotes the separation between transfer resonators $b_{2}$ and $c_{1}$. The speed of light in the waveguide is defined as $v$.

In Eq.~\eqref{eq:H_I} we have made the simplification that all transfer resonators have the same frequency $\omega_{0}$. This assumption is justified for several reasons. First, the transfer resonators are over-coupled to the waveguide, with linewidths of $\gamma_{\mu j}/2\pi\sim 10-20$ MHz assumed in this work. Second, the frequency of the parametric drive that produces the beamsplitter interaction can be tuned to adjust the frequency of the photon that is emitted into the waveguide~\cite{Gheeraert2020}. Finally, flux-tunable qubits may be utilized as transfer resonators, as experimentally demonstrated in Ref.~\cite{Kannan2023} (the state-transfer protocol discussed here is indifferent to whether the transfer resonators are qubits or are harmonic oscillators, as they are ideally not excited beyond the first-excited state).

In the following we take $d_{b}=d_{c}=\lambda/4$~\cite{Gheeraert2020, Guimond2020}, where $\lambda=2\pi v / \omega_{0}$ is the wavelength of the photon emitted into the waveguide. We also assume symmetric decay rates $\gamma_{\mu j}=\gamma, \; \forall \; \mu, j$. We explore the effects of asymmetric decay rates in Appendix~\ref{app:input_output}. The authors of Ref.~\cite{Gheeraert2020} have analyzed the case of deviations from $d=\lambda/4$, finding robustness of the directionality properties to deviations from the ideal value.

As detailed in Refs.~\cite{Guimond2020, Lalumiere2013, Gheeraert2020}, the waveguide can be eliminated in favor of an effective master equation description in terms of coupled transfer resonators
\begin{align}
\label{eq:master_2}
\frac{d\rho}{dt} = -i\left[ \hat{H}_{\mathrm{eff}},\rho \right] +\mathcal{D}[\hat{L}_{\rm R}^{(b)}(\omega_{0})+\hat{L}_{\rm R}^{(c)}(\omega_{0})]\rho \\ \nonumber + \mathcal{D}[\hat{L}_{\mathrm{L}}^{(b)}(\omega_{0})+\hat{L}_{\mathrm{L}}^{(c)}(\omega_{0})]\rho,
\end{align}
where 
\begin{align}
\label{eq:H_eff}
&\hat{H}_{\mathrm{eff}} = \hat{H}_{\mathrm{sys}}(t) + \sum_{\mu=b,c}\gamma (\hat{\mu}_{1}^{\dagger}\hat{\mu}_{2} + \hat{\mu}_{2}^{\dagger}\hat{\mu}_{1}) \\ \nonumber 
 &-\frac{i}{2}\Big( [\hat{L}_{\rm R}^{(c)}(\omega_{0})]^{\dagger}\hat{L}_{\rm R}^{(b)}(\omega_{0}) + [\hat{L}_{\mathrm{L}}^{(b)}(\omega_{0})]^{\dagger}\hat{L}_{\mathrm{L}}^{(c)}(\omega_{0}) - \mathrm{H. c.} \Big),
\end{align}
and $\mathcal{D}[\hat{\mathcal{O}}]\rho$ is the standard notation for the dissipator associated with collapse operator $\hat{\mathcal{O}}$ applied to the density matrix $\rho$. 
The second term in the first line of Eq.~\eqref{eq:H_eff} is the aforementioned effective exchange interaction between transfer resonators within the same GUE that is mediated by the waveguide. Choosing $J=-\gamma$ for the static coupling strength cancels this unwanted interaction. 
This effective description assumes that the GUEs exchange real photons only at the resonance frequency $\omega_{0}$, while only virtual photons are exchanged at other frequencies~\cite{Lalumiere2013}. 

Now we consider the situation where no photons escape either to the left or to the right, known as the {\it dark-state condition}~\cite{Cirac1997, Guimond2020, Stannigel2011}. Using the language of quantum trajectories, a pure state $|\psi(t)\rangle$ then evolves under the non-hermitian Hamiltonian \footnote{From the perspective of a dark state, this Hamiltonian actually is Hermitian.}
\begin{align}
\label{eq:H_nh}
\hat{H}_{\mathrm{eff}}^{\mathrm{nh}} = \hat{H}_{\mathrm{eff}} - \frac{i}{2}\hat{L}_{\rm R}^{\dagger}\hat{L}_{\rm R}
- \frac{i}{2}\hat{L}_{\mathrm{L}}^{\dagger}\hat{L}_{\mathrm{L}},
\end{align}
where we have defined the collective decay operators $\hat{L}_{\rm L/R} = \hat{L}_{\rm L/R}^{(b)}(\omega_{0})+\hat{L}_{\rm L/R}^{(c)}(\omega_{0})$. The non-hermitian terms in Eq.~\eqref{eq:H_nh} interfere with those in $\hat{H}_{\mathrm{eff}}$. After inserting the definitions \eqref{eq:L_defs} and noting that $\omega_{0}d_{\mu}/v=\pi/2, \mu=b,c,$ we obtain
\begin{align}
\label{eq:H_nh2}
\hat{H}_{\mathrm{eff}}^{\mathrm{nh}} &= 
\sum_{\mu=b,c}[g_{\mu}(t)(\hat{\mu}_{1}^{\dagger}\hat{\mu}_{3} + \hat{\mu}_{2}^{\dagger}\hat{\mu}_{4}) + \mathrm{H. c.}] \\ \nonumber 
&\quad -\frac{i}{2}\sum_{\substack{d=\mathrm{R,L} \\ \mu=b,c}}[\hat{L}_{d}^{(\mu)}(\omega_{0})]^{\dagger}\hat{L}_{d}^{(\mu)}(\omega_{0}) \\ \nonumber 
&\quad -i\Big([\hat{L}_{\rm R}^{(c)}(\omega_{0})]^{\dagger}\hat{L}_{\rm R}^{(b)}(\omega_{0}) + [\hat{L}_{\mathrm{L}}^{(b)}(\omega_{0})]^{\dagger}\hat{L}_{\mathrm{L}}^{(c)}(\omega_{0})\Big)
\end{align}
The structure of the last line of Eq.~\eqref{eq:H_nh2} encodes the directionality: a left-propagating state $|\psi_{\mathrm{L}}\rangle$ initialized in GUE $c$ only couples to a left-propagating state in GUE $b$, and a right-propagating state $|\psi_{\rm R}\rangle$ in GUE $b$ couples only to a right-propagating state in GUE $c$. That is to say, interference between terms in $H_{\rm eff}$ and those arising from the dissipators leads to the cancellation of terms that would allow for e.g. the creation of the state $|\psi_{\rm L}\rangle$ in GUE $c$ along with the annihilation of the state $|\psi_{\rm L}\rangle$ in GUE $b$.

We now proceed to derive the necessary control pulses $g_{\mu}(t), \mu=b,c$ to perform state transfer between GUEs $b$ and $c$. Without loss of generality, we consider the state transfer problem beginning with the initial state $|\psi_{i}\rangle=|\psi_{\rm R}\rangle|00\rangle|00\rangle|00\rangle$ and ending with the final state $|\psi_{f}\rangle=|00\rangle|00\rangle|00\rangle|\psi_{\rm R}\rangle$ after some specified time. We have ordered the kets as $b_{3}, b_{4}, b_{1}, b_{2}, c_{1}, c_{2}, c_{3}, c_{4}$, tracking the progress of population moving from GUE $b$ to GUE $c$. The state of the system at intermediate times is
\begin{align}
\label{eq:psi_GUE}
|\psi(t)\rangle &= \alpha_{Rb}(t)|\psi_{R b}\rangle + \alpha_{Rbt}(t)|\psi_{R b t}\rangle \\ \nonumber  &\quad+ \alpha_{Rct}(t)|\psi_{R c t}\rangle + \alpha_{Rc}(t)|\psi_{R c}\rangle,
\end{align}
where $|\psi_{R\mu}\rangle$ ($|\psi_{R\mu t}\rangle$) are the states where the data cavities (transfer resonators) of GUE $\mu$ are occupied. The time evolution of this state is governed by the time-dependent Schr\"odinger equation $i\frac{d}{dt}|\psi(t)\rangle=\hat{H}_{\mathrm{eff}}^{\mathrm{nh}}|\psi(t)\rangle$. Inserting Eq.~\eqref{eq:psi_GUE} into the Schr\"odinger equation yields the following four coupled differential equations for the coefficients
\begin{align}
\label{eq:diffeq}
i\dot\alpha_{Rb}(t) &= g_{b}^{*}(t)\alpha_{Rbt}(t) \\ \nonumber 
i\dot\alpha_{Rbt}(t) &= -i\gamma\alpha_{Rbt}(t) + g_{b}(t)\alpha_{Rb}(t) \\ \nonumber 
i\dot\alpha_{Rct}(t) &= -i\gamma\alpha_{Rct}(t) + g_{c}(t)\alpha_{Rc}(t) + 2\gamma e^{i\phi}\alpha_{Rbt}(t) \\ \nonumber 
i\dot\alpha_{Rc}(t) &= g_{c}^{*}(t)\alpha_{Rct}(t),
\end{align}
defining $\phi=\omega_{0}d_{bc}/v$.
Analytically solving for the pulses $g_{b}(t), g_{c}(t)$ that satisfy these four differential equations subject to the initial and final conditions does not appear to be straightforward. To proceed, we adiabatically eliminate the transfer resonators. This approximation is motivated by the transfer resonators' overcoupling to the waveguide, causing any population to immediately be emitted. That is, the adiabaticitiy condition is $g_{\mu}\ll \gamma$. Setting $\dot\alpha_{Rbt}(t)\approx0, \dot\alpha_{Rct}(t)\approx0$, we obtain
\begin{align}
\dot\alpha_{Rb}(t) &=-\frac{|g_{b}(t)|^2}{\gamma}\alpha_{Rb}(t) \\ \nonumber 
\dot\alpha_{Rc}(t) &= -\frac{|g_{c}(t)|^2}{\gamma}\alpha_{Rc}(t)+2ie^{i\phi}\frac{g_{c}^{*}(t)g_{b}(t)}{\gamma}\alpha_{Rb}(t),
\end{align}
describing an effective directional interaction between the two GUEs. Choosing $\phi=\pi/2$ (which amounts to a specific spacing of the GUEs) yields the exact same differential equation for directional state transfer as in Ref.~\cite{Stannigel2011}, c.f. Eq. (47). There, Stannigel {\it et al.} studied state transfer between two qubits along a 1D waveguide in an optomechanical setting.
It is remarkable that we recover the results of Ref.~\cite{Stannigel2011}, given that we made no assumption about directionality of the waveguide itself or preferential coupling to left- or right-propagating modes. Instead, directionality here is due to appropriate spacing of the transfer resonators along the waveguide.

It is worth emphasizing here that the state transfer protocol trivially extends to the case of superposition states. We say it is trivial because the Hamiltonian~\eqref{eq:H_nh2} is number conserving: thus it immediately follows that we can execute the state transfer $(\beta_{1}|00\rangle+\beta_{2}|\psi_{\rm R}\rangle)|00\rangle|00\rangle|00\rangle \rightarrow |00\rangle|00\rangle|00\rangle(\beta_{1}|00\rangle+\beta_{2}|\psi_{\rm R}\rangle)$ if we can perform $|\psi_{Rb}\rangle\rightarrow|\psi_{Rc}\rangle$. 

We apply the results of Ref.~\cite{Stannigel2011} to obtain pulses that yield a time-symmetric emitted wave packet, allowing for it to be absorbed using a time-reversed pulse. One set of solutions is
\begin{align}
\label{eq:gb}
g_{b}(t) &= \frac{\sqrt{\frac{\gamma}{2}}\exp(-\zeta t^2/2)}{\sqrt{\frac{1}{\xi}-\sqrt{\frac{\pi}{4\zeta}}\erf(\sqrt{\zeta}t)}}, \\ \nonumber 
g_{c}(t) &= g_{b}(-t),
\end{align}
where $\xi, \zeta$ must be adjusted appropriately to satisfy the boundary conditions
\begin{align}
\alpha_{Rb}(t_{i}) = |\alpha_{Rc}(t_{f})|=1, \qquad \alpha_{Rb}(t_{f})=\alpha_{Rc}(t_{i})=0,
\end{align}
where $t_{i}, t_{f}$ are the initial and final times, respectively and $t_{i}=-t_{f}$. We find empirically that setting $\zeta\lesssim \xi^2 \pi / 4$ yields good results, leaving $\xi$ the only variable to tune. Throughout this work we set $\xi/2\pi=0.95$ MHz and $\gamma/2\pi = 20$ MHz. Moreover we find that optimization over scale factors $\lambda_{\mu}$, where $g_{\mu}(t)\rightarrow\lambda_{\mu}g_{\mu}(t), \mu=b,c$ improves state-transfer fidelities (discussed below). We utilize $\lambda_{b}=1.018,\lambda_{c}=1.017$ in the remainder of this work.

\begin{figure}
    \centering
    \includegraphics[width=\columnwidth]{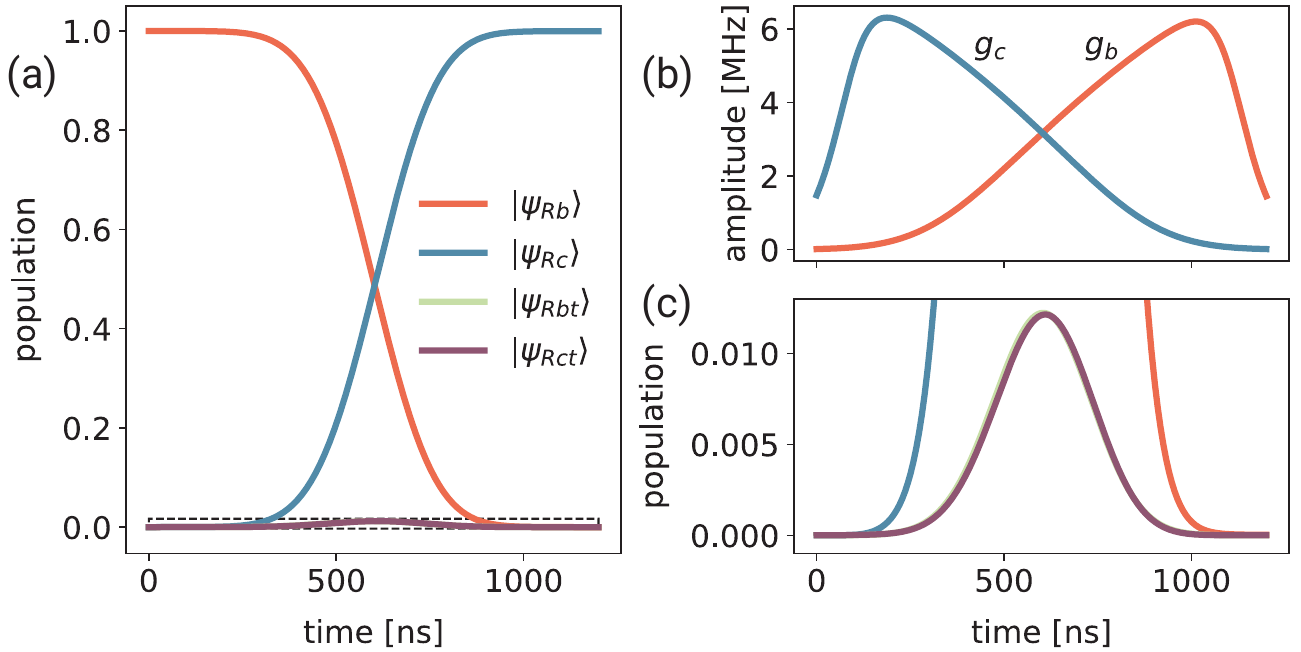}
    \caption{\label{fig:GUE_state_transfer} State-transfer protocol between two GUEs. (a) Population swaps from GUE $b$ to GUE $c$ due to the control pulses (b) applied between the GUEs and the transfer resonators. (c) The transfer resonators are minimally occupied during the course of the operation [the area plotted in (c) is highlighted by a dashed box in (a)].}
\end{figure}

The dark-state condition $\hat{L}_{\mathrm{L}}|\psi(t)\rangle=0$ is automatically true for the case of rightward-state-transfer considered here. However, the condition $\hat{L}_{\rm R}|\psi(t)\rangle=\sqrt{\gamma}[\alpha_{Rbt}(t)-ie^{-i\phi}\alpha_{Rct}(t)]=0\;\forall \; t,$ is only approximately satisfied due to suppressed (by the adiabatic condition) but nonzero occupation of the transfer resonators during the state-transfer protocol [see Fig.~\ref{fig:GUE_state_transfer}]. Violation of the dark-state condition represents population lost to the waveguide and limits state-transfer fidelities as we show below.

Towards performing numerical simulations of the full state-transfer protocol, we now include a third GUE (labeled $a$) to the left of GUE $b$, see Fig.~\ref{fig:fig10}. The master equation is
\begin{align}
\nonumber
\frac{d\rho}{dt} &= -i\left[ \hat{H}_{\mathrm{eff}},\rho \right] +\mathcal{D}[\hat{L}_{\rm R}^{(a)}(\omega_{0})+\hat{L}_{\rm R}^{(b)}(\omega_{0})+\hat{L}_{\rm R}^{(c)}(\omega_{0})]\rho \\  
\label{eq:mastereq_3}
&\quad+ \mathcal{D}[\hat{L}_{\mathrm{L}}^{(a)}(\omega_{0})+\hat{L}_{\mathrm{L}}^{(b)}(\omega_{0})+\hat{L}_{\mathrm{L}}^{(c)}(\omega_{0})]\rho,
\end{align}
where
\begin{align}
\label{eq:H_eff_3}
&\hat{H}_{\mathrm{eff}} = \hat{H}_{\mathrm{sys}}(t) + \sum_{\mu=a,b,c}\gamma (\hat{\mu}_{1}^{\dagger}\hat{\mu}_{2} + \hat{\mu}_{2}^{\dagger}\hat{\mu}_{1}) \\ \nonumber 
 &-\frac{i}{2}\Big([\hat{L}_{\mathrm{R}}^{(c)}(\omega_{0})]^{\dagger} 
\hat{L}_{\mathrm{R}}^{(b)}(\omega_{0})
+[\hat{L}_{\mathrm{L}}^{(b)}(\omega_{0})]^{\dagger}
\hat{L}_{\mathrm{L}}^{(c)}(\omega_{0}) \\ \nonumber 
&\quad+[\hat{L}_{\mathrm{R}}^{(c)}(\omega_{0})]^{\dagger}
\hat{L}_{\mathrm{R}}^{(a)}(\omega_{0})
+[\hat{L}_{\mathrm{L}}^{(a)}(\omega_{0})]^{\dagger}
\hat{L}_{\mathrm{L}}^{(c)}(\omega_{0}) \\ \nonumber 
&\quad +[\hat{L}_{\mathrm{R}}^{(b)}(\omega_{0})]^{\dagger}
\hat{L}_{\mathrm{R}}^{(a)}(\omega_{0})
+[\hat{L}_{\mathrm{L}}^{(a)}(\omega_{0})]^{\dagger}
\hat{L}_{\mathrm{L}}^{(b)}(\omega_{0}) - \mathrm{H. c.} \Big),
\end{align}
and $\hat{H}_{\mathrm{sys}}(t)$ is as in Eq.~\eqref{eq:Hsys} with the sum extended to include GUE $a$. As before, we obtain the non-hermitian effective Hamiltonian by assuming the dark-state condition, yielding~\cite{Stannigel2011, Guimond2020}
\begin{align}
\label{eq:Hnh_3}
\hat{H}_{\mathrm{eff}}^{\mathrm{nh}} &= 
\sum_{\mu=a,b,c}[g_{\mu}(t)(\hat{\mu}_{1}^{\dagger}\hat{\mu}_{3} + \hat{\mu}_{2}^{\dagger}\hat{\mu}_{4}) + \mathrm{H. c.}] \\ \nonumber 
&\quad-\frac{i}{2}\sum_{\substack{d=\mathrm{R,L} \\ \mu=a,b,c}}[\hat{L}_{d}^{(\mu)}(\omega_{0})]^{\dagger}\hat{L}_{d}^{(\mu)}(\omega_{0})
\\ \nonumber 
&\quad-i\Big(
[\hat{L}_{\mathrm{R}}^{(c)}(\omega_{0})]^{\dagger} 
\hat{L}_{\mathrm{R}}^{(b)}(\omega_{0})
+[\hat{L}_{\mathrm{L}}^{(b)}(\omega_{0})]^{\dagger}
\hat{L}_{\mathrm{L}}^{(c)}(\omega_{0}) \\ \nonumber 
&\quad+[\hat{L}_{\mathrm{R}}^{(c)}(\omega_{0})]^{\dagger}
\hat{L}_{\mathrm{R}}^{(a)}(\omega_{0})
+[\hat{L}_{\mathrm{L}}^{(a)}(\omega_{0})]^{\dagger}
\hat{L}_{\mathrm{L}}^{(c)}(\omega_{0}) \\ \nonumber 
&\quad +[\hat{L}_{\mathrm{R}}^{(b)}(\omega_{0})]^{\dagger}
\hat{L}_{\mathrm{R}}^{(a)}(\omega_{0})
+[\hat{L}_{\mathrm{L}}^{(a)}(\omega_{0})]^{\dagger}
\hat{L}_{\mathrm{L}}^{(b)}(\omega_{0})\Big),
\end{align}
a relatively straightforward generalization of Eqs.~\eqref{eq:H_nh}-\eqref{eq:H_nh2}. 
We define the decay operators now utilizing $a_{1}$ as the origin, thus we have e.g. $\hat{L}_{\rm R}^{(a)}(\omega_{0})=\sqrt{\gamma_{a1}}\hat{a}_{1}-i\sqrt{\gamma_{a2}}\hat{a}_{2}$ and $\hat{L}_{\rm R}^{(c)}(\omega_{0})=(-i)^2e^{-i(\phi_{ab}+\phi_{bc})}(\sqrt{\gamma_{c1}}\hat{c}_{1}-i\sqrt{\gamma_{c2}}\hat{c}_{2})$, where $\phi_{ab}=\omega_{0}d_{ab}/v, \phi_{bc}=\omega_{0}d_{bc}/v$.
We take $g_{a}(t)=g_{c}(t)$ to simultaneously catch the emitted wave packets in both receiver GUEs.
In the single-rail case the initial basis states are $\{|\overline{0}\rangle\equiv|00\rangle|00\rangle|00\rangle,|\overline{\psi_{\rm R}}\rangle\equiv|00\rangle|\psi_{\rm R}\rangle|00\rangle,|\overline{\psi_{\rm L}}\rangle\equiv|00\rangle|\psi_{\rm L}\rangle|00\rangle\},$ (ordering the kets as $a_{3}, a_{4}, b_{3}, b_{4}, c_{3},c_{4}$ and omitting the transfer resonators), while in the dual-rail case they are $\{ |\overline{\psi_{\rm R}}\rangle|\overline{0}\rangle,|\overline{\psi_{\rm L}}\rangle|\overline{0}\rangle,|\overline{0}\rangle|\overline{\psi_{\rm R}}\rangle,|\overline{0}\rangle|\overline{\psi_{\rm L}}\rangle \}$. The final basis states are similarly defined.

Simulating the full state-transfer protocol is challenging even in the single-rail case due to the 12 subsystems involved (6 data cavities and 6 transfer resonators). We take advantage of the fact that the Hamiltonian is number conserving, where the only non-number conserving processes are due to decay or heating. Thus we specify a global excitation number cutoff when constructing our basis, as opposed to including e.g. $3^{12}$ basis states (allowing for two excitations per mode to include heating effects)~\cite{Weiss2021, Zhang2010}. We take care to ensure that our results do not depend on the global excitation-number cutoff.

In the dual-rail case, we reconstruct the time evolution of a dual-rail state by appropriately tensoring together the time evolution of single-rail states \footnote{This assumes truly independent time evolution, e.g. if the different rails are connected to different waveguides such that the state transfer can be done in parallel. In the case where they are connected to the same waveguide, the state transfer would need to be accomplished serially.}. State-transfer is now performed pairwise, with the right GUE in the sender dual rail transmitting to the right receiver GUE and likewise for the left GUEs. On the one hand if these different pairs of GUEs are coupled to separate waveguides (in the experiment of Ref.~\cite{Qiu2023} the authors coupled two qubits via a 64 m Al cable where the cable left the 2D chip, suggesting an architecture where multiple cables can be braided in 3D), then the state-transfer problems proceeds as in the single-rail case. On the other hand, if all GUEs are connected to the same waveguide, each state-transfer operation includes passing through an ``inactive" GUE, see Fig.~\ref{fig:3_bit_GUE}. This pass-through GUE is inactive in the sense that the beamsplitter coupling between the data cavities and transfer resonators is turned off, however the transfer resonators are still coupled to the waveguide. Perhaps surprisingly, as discussed in Ref.~\cite{Gheeraert2020} and Appendix~\ref{app:input_output}, in the ideal case of parameter symmetry this GUE serves only to impart a Wigner time delay on the transmitted photon~\cite{Wigner1955, Hauge1989, Gheeraert2020}. Thus the previously derived state-transfer protocol can be applied to this case, provided the retarded time of the receiver GUE(s) are redefined to account for the Wigner delay (in addition to the delay due to the finite propogation time of the photon). We discuss the non-ideal case of parameter asymmetry in Appendix ~\ref{app:input_output}.

The average state-transfer fidelity $F_{\mathrm{st}}$ is calculated by averaging the individual state-transfer fidelities 
\begin{align}
F_{\mathrm{st}, \mu}=\langle \mu| \Tr_{1}[\mathcal{E}_{\mathrm{st}}(|\mu\rangle\langle\mu|)]|\mu\rangle
\end{align}
over the initial basis states as well as their $\textit{X}$ and $\textit{Y}$ superpositions. The trace is performed over the transfer resonators as well as the initial data cavities. In the single-rail case, $\mathcal{E}_{\mathrm{st}}$ includes only time evolution under the state-transfer protocol, while for dual rail we also include a projective measurement onto the dual-rail states. Again, we include this measurement only to obtain an estimate for the post-selected fidelity of a dual-rail QRAM query, and emphasize that such a measurement cannot actually be performed during a query. We include the effects of nonradiative decay of the transfer resonator as well as transfer-resonator dephasing, see Tab.~\ref{tab:gue_fidel} (in addition to the data-cavity coherence times in Tab.~\ref{tab:params}). 
\begin{figure}
    \centering
    \includegraphics[width=\columnwidth]{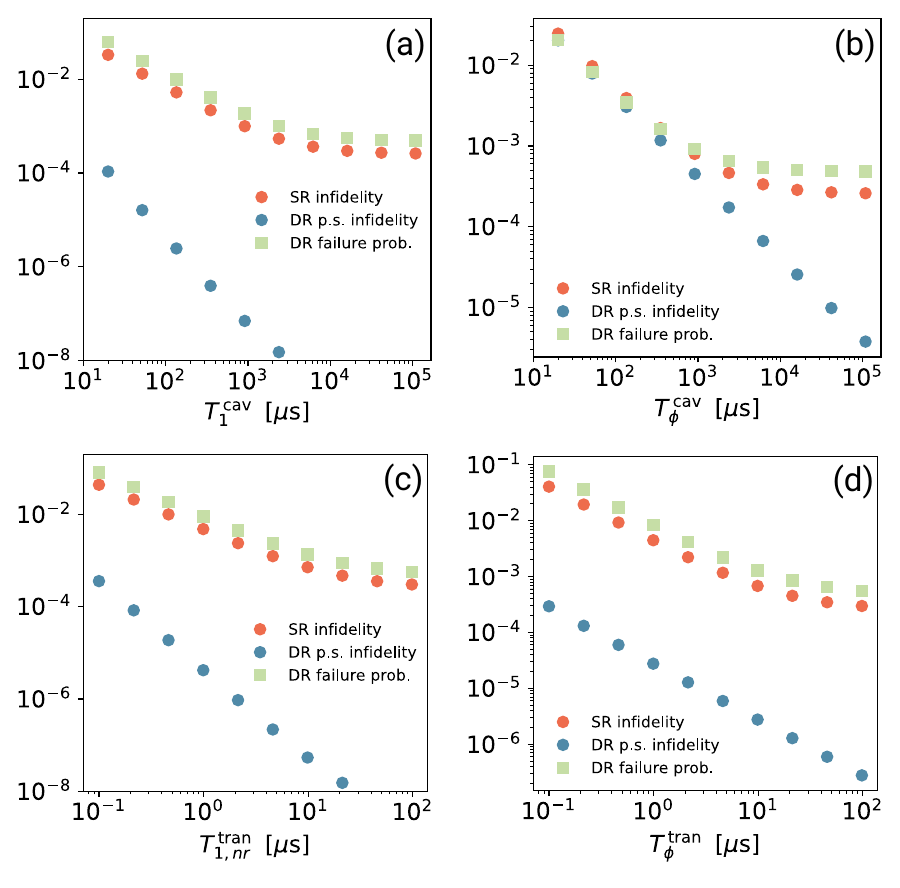}
    \caption{\label{fig:GUE_fidelity_T1_sweep} State-transfer infidelity and failure probability. We sweep over data cavity (a) $T_{1}^{\mathrm{cav}}$  and (b) $T_{\phi}^{\mathrm{cav}}$ as well as (c) $T_{1, \mathrm{nr}}^{\mathrm{tran}}$ and (d) $T_{\phi}^{\mathrm{tran}}$. The plateau of the single-rail infidelity (and failure probability) is due to the inherent infidelity associated with our state-transfer protocol, at the level of $10^{-4}$ for our parameters. In the dual-rail case decay to the vacuum state is detectable, thus the post-selected infidelity scales as $(T_{1}^{\mathrm{cav}})^{-2}, (T_{1, \mathrm{nr}}^{\mathrm{tran}})^{-2}$.  }
\end{figure}

In the single-rail case and utilizing PS2 estimates, we obtain a state-transfer infidelity of $3.3\times10^{-4}$, see Tab.~\ref{tab:gue_fidel}. Decay to the vacuum state generally limits the fidelity of the state-transfer protocol, due both to decoherence as well as to violation of the dark-state condition. 
Thus upon sweeping the coherence times of the data cavities and transfer resonators, the single-rail infidelity plateaus at the $10^{-4}$ level [see Fig.~\ref{fig:GUE_fidelity_T1_sweep}]. 
Utilizing a dual-rail architecture [see Fig.~\ref{fig:3_bit_GUE}] helps to mitigate this issue, where now decay to the vacuum is detectable. Therefore the infidelity of the dual-rail state-transfer protocol scales quadratically with cavity decay and transfer-resonator nonradiative decay [see Fig.~\ref{fig:GUE_fidelity_T1_sweep}].
In terms of PS2 estimates, the post-selected state-transfer infidelity is $4.0\times10^{-6}$ [see Tab.~\ref{tab:gue_fidel}]. The failure rate in this case is $6.1\times10^{-4}$, which is as expected roughly twice the single-rail infidelity.

\section{Non-idealities in GUE state transfer}
\label{app:input_output}

Various non-idealities affect the ability to perform high-fidelity state transfer. These include (i) imperfect cancellation of the direct coupling between transfer resonators, (ii) detuning of the transfer resonators from resonance, (iii) deviation from $d=\lambda/4$ of the distance between the transfer resonators, and (iv) disorder in the radiative decay rates of the transfer resonators. Imperfect cancellation (i) can be addressed by using tunable beamsplitter interactions (as discussed in the main text). Nonidealities (ii-iii) can be addressed by utilizing e.g. flux-tunable transmons as the transfer resonators, as in the experiment in Ref.~\cite{Kannan2023}. This allows for the emitters to be tuned into resonance at the specific frequency appropriate for their true separation along the waveguide. Disorder in the decay rates (iv) does not appear to be easily tunable in situ, thus we focus on characterizing the directionality properties associated with asymmetric decay rates. 

There are three processes that could be affected: emission, absorption and ``pass through." Note that due to time-reversal symmetry, we need only consider one of emission or absorption. The directionality properties for the other process immediately follows. With respect to the pass-through problem, in the dual-rail architecture both rails may be connected to the same waveguide. Thus it becomes necessary for photons to pass through an inactive GUE. In the case of symmetric decay rates, wave packets pass through undistorted, incuring only a Wigner time delay~\cite{Gheeraert2020, Wigner1955}. Disorder in the decay rates of the inactive GUE could result in reflection or distortion of the transmitted signal, and we analyze the symmetric and asymmetric cases below.

These problems are most appropriately analyzed in the context of input-output theory~\cite{Gardiner1985}, considering a wave packet incident on a GUE. The derivation of the associated Langevin equations can be found in e.g. Refs.~\cite{Gheeraert2020, Guimond2020}, and we obtain
\begin{align}
\label{eq:Langevin}
\frac{d \hat{b}_{1}(t)}{dt} &= -\gamma_{1}\hat{b}_{1}(t)
-ig_{b}(t)\hat{b}_{3}(t) \\ \nonumber
&\quad -\sqrt{\gamma_{1}}e^{-i\omega_{0}d/v}\hat{a}_{\rm R}^{\mathrm{in}}(t)-\sqrt{\gamma_{1}}\hat{a}_{\rm L}^{\mathrm{in}}(t), \\ \nonumber 
\frac{d \hat{b}_{2}(t)}{dt} &= -\gamma_{2}\hat{b}_{2}(t)-ig_{b}(t)\hat{b}_{4}(t) \\ 
&\quad-\sqrt{\gamma_{2}}e^{-i\omega_{0}d/v}\hat{a}_{\rm L}^{\mathrm{in}}(t)-\sqrt{\gamma_{2}}\hat{a}_{\rm R}^{\mathrm{in}}(t), \\ 
\frac{d \hat{b}_{3}(t)}{dt} &= -ig_{b}(t)\hat{b}_{1}(t), \\
\label{eq:Langevin_4}
\frac{d \hat{b}_{4}(t)}{dt} &= -ig_{b}(t)\hat{b}_{2}(t),
\end{align}
in terms of the input fields $\hat{a}_{\rm L,R}^{\rm in}(t)$ and
\begin{align}
\label{eq:Langevin_out}
\frac{d \hat{b}_{1}(t)}{dt} &= \gamma_{1}\hat{b}_{1}(t) -ig_{b}(t)\hat{b}_{3}(t)
\\
\label{eq:Langevin_out_2}
&\quad -\sqrt{\gamma_{1}}e^{-i\omega_{0}d/v}\hat{a}_{\rm R}^{\mathrm{out}}(t)-\sqrt{\gamma_{1}}\hat{a}_{\rm L}^{\mathrm{out}}(t), \\ \nonumber 
\frac{d \hat{b}_{2}(t)}{dt} &= \gamma_{2}\hat{b}_{2}(t) -ig_{b}(t)\hat{b}_{4}(t)
\\ \nonumber &\quad-\sqrt{\gamma_{2}}e^{-i\omega_{0}d/v}\hat{a}_{\rm L}^{\mathrm{out}}(t)-\sqrt{\gamma_{2}}\hat{a}_{\rm R}^{\mathrm{out}}(t),
\end{align}
in terms of the output fields $\hat{a}_{\rm L,R}^{\rm out}(t)$.
Here, $d$ is the distance between the GUEs, $\hat{b}_{i}$ is the annihilation operator for transfer resonator $i$, 
and $\hat{a}_{d}^{\mathrm{in/out}}$ are the input/output fields in direction $d$.  
We first discuss the pass-through protocol, before returning to absorption. 

\subsection{Pass through}
\subsubsection{Symmetric decay rates}

In this case we take $\gamma_{1}=\gamma_{2}=\gamma$. In the pass-through protocol the beamsplitter interactions are turned off, $g_{b}(t)=g_{c}(t)=0$. We thus obtain the particularly simple Langevin equations for the right and left collective modes
\begin{align}
\label{eq:a_b_in}
\frac{d \hat{b}_{\rm R, L}(t)}{dt} = -2\sqrt{\gamma}\hat{a}_{\rm R, L}^{\mathrm{in}}(t)-\gamma \hat{b}_{\rm R, L}(t) \\ 
\label{eq:a_b_out}
= -2\sqrt{\gamma}\hat{a}_{\rm R, L}^{\mathrm{out}}(t)+\gamma \hat{b}_{\rm R, L}(t),
\end{align}
$\hat{b}_{\rm R}(t) = i \hat{b}_{1}(t) + \hat{b}_{2}(t)$ and $\hat{b}_{\rm L}(t) = \hat{b}_{1}(t) + i\hat{b}_{2}(t)$. We read off the input-output relations
\begin{align}
\hat{a}_{\rm R, L}^{\mathrm{out}}(t) = \hat{a}_{\rm R,L}^{\mathrm{in}}(t) + \sqrt{\gamma}\hat{b}_{\rm R, L}(t).
\end{align}
In this ideal case, the right and left modes completely decouple.
We can solve for the output fields in terms of the input fields by working in Fourier space, defining
\begin{align}
\hat{a}_{\rm R, L}^{\mathrm{in,\, out}}(t) = \int_{-\infty}^{\infty}\frac{d\omega}{\sqrt{2\pi}}\hat{a}_{\rm R, L}^{\mathrm{in,\, out}}(\omega)e^{-i\omega t},
\end{align}
with similar definitions for the fields $\hat{b}_{\rm R,L}(t)$. We emphasize here that because we are working in the rotating frame, zero frequency indicates resonance with the emitter frequencies. Using Eq.~\eqref{eq:a_b_in} we obtain
\begin{align}
\hat{b}_{\rm R,L}(\omega) = \frac{2\sqrt{\gamma}}{i\omega-\gamma}\hat{a}_{\rm R,L}^{\mathrm{in}}(\omega),
\end{align}
leading immediately to
\begin{align}
\hat{a}_{\rm R,L}^{\mathrm{out}}(\omega) = \frac{\omega - i\gamma}{\omega + i\gamma}\hat{a}_{\rm R,L}^{\rm in}(\omega).
\end{align}
Thus we observe that the output fields are related to the input fields by a frequency-dependent prefactor that has unit modulus. Interpreting this prefactor as a phase $e^{-i\phi(\omega)}=\frac{\omega - i\gamma}{\omega + i\gamma}$, we obtain 
\begin{align}
\phi(\omega) = \arctan(\frac{2\gamma \omega}{\omega^{2}-\gamma^{2}}).
\end{align}
Taylor expanding about $\omega=0$ yields
\begin{align}
\phi(\omega) = -\frac{2\omega}{\gamma} + \frac{2\omega^3}{3\gamma^3} + \mathcal{O}\left(\left[\omega/\gamma\right]^5\right).
\end{align}
For input fields that are nearly resonant with the emitters such that the cubic and higher terms can be neglected, the overall effect is a Wigner time delay~\cite{Gheeraert2020,Hauge1989, Wigner1955}
\begin{align}
\hat{a}_{\rm R,L}^{\mathrm{out}}(t) = \hat{a}_{\rm R, L}^{\mathrm{in}}(t-2/\gamma). 
\end{align}
Thus the inactive GUE can be modeled as merely producing a phase shift, in addition to that produced by the time delay associated with the spatial separation between emitters. Thus all results derived for the GUE architecture (specifically those associated with state transfer between GUEs) are immediately applicable to the dual-rail GUE architecture, where the phase shift between GUEs acquires a contribution from the Wigner delay associated with traversing the pass-through GUE.

\subsubsection{Asymmetric decay rates}

\begin{figure}
    \centering
    \includegraphics[width=\columnwidth]{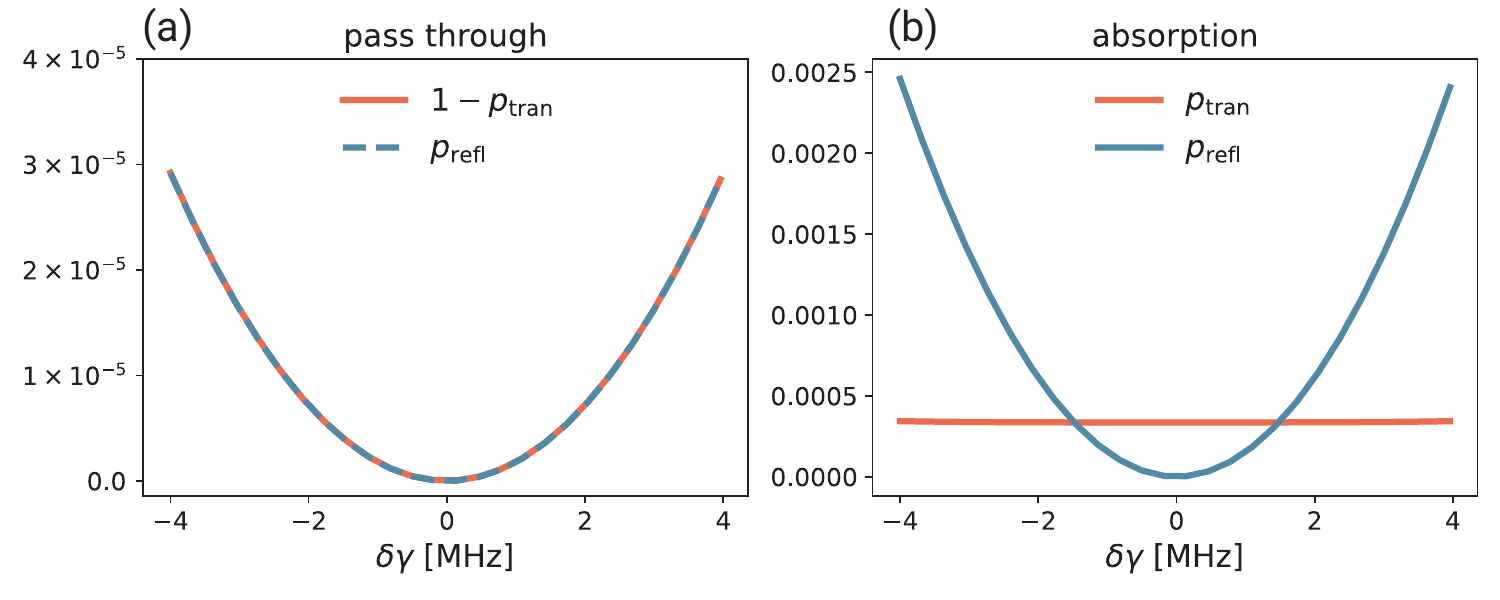}
    \caption{\label{fig:reflect_transmit} Reflection and transmission probabilities during pass through and absorption. (a) The pass-through protocol is relatively insensitive to decay-rate asymmetries. The reflection probability and deviation of the transmission probability from unity are both less than $0.005\%$ for $\delta\gamma/\gamma\leq 20\%$. (b) For the absorption process, the transmission probability is nearly constant as a function of $\delta\gamma$. This should ideally be zero, and is not due to slight violation of the dark-state condition. The reflection probability varies quadratically with $\delta\gamma$ [see Eqs.~\eqref{eq:aRout_dev}-\eqref{eq:prefl}] and remains below the transmission probability for $\delta\gamma/\gamma\lesssim 10\%$. }
    
\end{figure}

We now return to Eqs.~\eqref{eq:Langevin}-\eqref{eq:Langevin_out} and allow for $\gamma_{1}\neq\gamma_{2}$. The input-output relations are
\begin{align}
\label{eq:input_output_general_R}
\hat{a}_{\rm R}^{\mathrm{out}}(t) = \hat{a}_{\rm R}^{\mathrm{in}}(t) + i\sqrt{\gamma_{1}}\hat{b}_{1}(t)+\sqrt{\gamma_{2}}\hat{b}_{2}(t), \\
\label{eq:input_output_general_L}
\hat{a}_{\rm L}^{\mathrm{out}}(t) = \hat{a}_{\rm L}^{\mathrm{in}}(t) + \sqrt{\gamma_{1}}\hat{b}_{1}(t)+i\sqrt{\gamma_{2}}\hat{b}_{2}(t),
\end{align}
where Eqs.~\eqref{eq:input_output_general_R}-\eqref{eq:input_output_general_L} are not decoupled as they were in the symmetric case. 
Again working in Fourier space, we obtain
\begin{align}
\label{eq:inputoutputasymmetry}
\hat{a}_{\rm R,L}^{\mathrm{out}}(\omega)=\pm\frac{\omega(\gamma_{1}-\gamma_{2})}{(\omega+i\gamma_{1})(\omega+i\gamma_{2})}\hat{a}_{L\rm ,R}^{\mathrm{in}}(\omega) \\ \nonumber 
+\frac{\gamma_{1}\gamma_{2}+\omega^2}{(\omega+i\gamma_{1})(\omega+i\gamma_{2})}
\hat{a}_{\rm R,L}^{\mathrm{in}}(\omega).
\end{align}
Defining $\gamma=(\gamma_{1}+\gamma_{2})/2$ and $\delta\gamma = (\gamma_{1} - \gamma_{2})$ and expanding Eq.~\eqref{eq:inputoutputasymmetry} up to second order in $\delta\gamma$ yields
\begin{align}
\label{eq:aRout_dev}
\hat{a}_{\rm R,L}^{\mathrm{out}}(\omega)=\pm \frac{\omega \delta\gamma}{(\omega+i\gamma)^2}\hat{a}_{\rm L,R}^{\mathrm{in}}(\omega) \\ \nonumber 
+\left(\frac{\omega-i\gamma}{\omega+i\gamma} -\frac{\omega\delta\gamma^2}{2(\omega+i\gamma)^3}\right)\hat{a}_{\rm R,L}^{\mathrm{in}}(\omega) +\mathcal{O}(\delta\gamma^3).
\end{align}
Thus the leading-order effect of asymmetry in the decay rates is to cause reflection of the input waveform (as opposed to modifying the transmitted waveform shape or time delay, which is sub-leading order). To quantify the reduction in fidelity due to this reflection, we compute the probability of reflection
\begin{align}
\label{eq:prefl}
p_{\mathrm{refl}}=\frac{\int_{-\infty}^{\infty}d\omega 
\Big|\frac{\omega(\gamma_{1}-\gamma_{2})}{(\omega+i\gamma_{1})(\omega+i\gamma_{2})}
a_{\rm R}^{\mathrm{in}}(\omega)\Big|^2}{\int_{-\infty}^{\infty}d\omega|a_{\rm R}^{\mathrm{in}}(\omega)|^2},
\end{align}
assuming an input waveform traveling to the right and
utilizing the full formula Eq.~\eqref{eq:inputoutputasymmetry}. We obtain the input waveform from numerical solution of Eq.~\eqref{eq:diffeq}.
If we assume $\gamma/2\pi=20$ MHz and a decay-rate asymmetry of $10\%$ ($\delta\gamma/2\pi=2$ MHz), 
we obtain $p_{\mathrm{refl}}=7.2\times10^{-6}$ [see Fig.~\ref{fig:reflect_transmit}(a)]. Given that we only expect to achieve state-transfer infidelities on the order of $10^{-4}$, reflection off of a pass-through GUE due to decay-rate asymmetry is thus not a limiting factor. This robustness to decay-rate asymmetry is promising for a dual-rail architecture where all GUEs at one level of the tree are all connected to the same waveguide.

\subsection{Absorption}

We now assume time-dependent beamsplitter drives appropriate for absorption as in Eq.~\eqref{eq:gb}. These time-dependent drives complicate the analytics, and we proceed by numerically integrating Eqs.~\eqref{eq:Langevin}-\eqref{eq:Langevin_4}. The output fields are obtained from the input fields and the internal modes using the input-output relations Eqs.~\eqref{eq:input_output_general_R}-\eqref{eq:input_output_general_L}. With these solutions in hand we calculate the reflection and transmission probabilities in the case of a rightward-traveling input waveform
\begin{align}
p_{\rm refl} = \frac{\int_{-\infty}^{\infty}dt |a_{\rm L}^{\rm out}(t)|^2}{\int_{-\infty}^{\infty}dt |a_{\rm R}^{\rm in}(t)|^2}, \\
p_{\rm tran} = \frac{\int_{-\infty}^{\infty}dt |a_{\rm R}^{\rm out}(t)|^2}{\int_{-\infty}^{\infty}dt |a_{\rm R}^{\rm in}(t)|^2},
\end{align}
respectively. For perfect absorption, both of these probabilities should vanish. In the ideal case of $\gamma_{1}=\gamma_{2}$, the reflection probability indeed vanishes, see Fig.~\ref{fig:reflect_transmit}(b). This is consistent with the decoupling of the input-output relations Eqs.~\eqref{eq:input_output_general_R}-\eqref{eq:input_output_general_L} for $\gamma_{1}=\gamma_{2}$ (these equations are unchanged by the beamsplitter drive). The transmission probability is nonvanishing even for $\gamma_{1}=\gamma_{2}$ [see Fig.~\ref{fig:reflect_transmit}(b)], due to violation of the dark-state condition. This limits the fidelity of the state-transfer protocol in the symmetric case $\delta\gamma=0$ and represents population lost to the waveguide. The transmission probability is essentially constant, and thus for decay-rate variations of up to $\sim20\%$, the dark-state violation effect is the leading contributor to infidelity. We have also calculated (not shown) the coherent loss in fidelity due to $\delta\gamma\neq0$ by simulating the state-transfer protocol using the master equations ~\eqref{eq:master_2},~\eqref{eq:mastereq_3}. We find that for modest values of asymmetry $\delta\gamma/\gamma <0.2$, the violation of the dark-state condition still dominates the contribution to infidelity. It is thus interesting to explore in future work protocols for state-transfer that respect the dark-state condition, which would improve fidelities.

\bibliography{bib}
\end{document}